\documentclass[twocolumn]{aastex63}

\usepackage{amsmath}

\shorttitle{Inclination-dependent SFHs}
\shortauthors{Doore et al.}

\begin{document}

\title{On the Impact of Inclination-Dependent Attenuation on Derived Star Formation Histories: Results from Disk Galaxies in the Great Observatories Origins Deep Survey Fields}

\correspondingauthor{Keith Doore}
\email{kjdoore@uark.edu}

\author[0000-0001-5035-4016]{Keith~Doore}
\affiliation{Department of Physics, University of Arkansas, 226 Physics Building, 825 West Dickson Street, Fayetteville, AR 72701, USA}

\author[0000-0002-2987-1796]{Rafael~T.~Eufrasio}
\affiliation{Department of Physics, University of Arkansas, 226 Physics Building, 825 West Dickson Street, Fayetteville, AR 72701, USA}

\author[0000-0003-2192-3296]{Bret~D.~Lehmer}
\affil{Department of Physics, University of Arkansas, 226 Physics Building, 825 West Dickson Street, Fayetteville, AR 72701, USA}

\author[0000-0001-8473-5140]{Erik~B.~Monson}
\affiliation{Department of Physics, University of Arkansas, 226 Physics Building, 825 West Dickson Street, Fayetteville, AR 72701, USA}

\author{Antara~Basu-Zych}
\affiliation{NASA Goddard Space Flight Center, Code 662, Greenbelt, MD 20771, USA}
\affiliation{Center for Space Science and Technology, University of Maryland Baltimore County, 1000 Hilltop Circle, Baltimore, MD 21250, USA}

\author[0000-0002-9202-8689]{Kristen~Garofali}
\affiliation{NASA Goddard Space Flight Center, Code 662, Greenbelt, MD 20771, USA}

\author[0000-0001-5655-1440]{Andrew~Ptak}
\affiliation{NASA Goddard Space Flight Center, Code 662, Greenbelt, MD 20771, USA}
\affiliation{The Johns Hopkins University, Homewood Campus, Baltimore, MD 21218, USA}



\begin{abstract}

We develop and implement an inclination-dependent attenuation prescription for spectral energy distribution (SED) fitting and study its impact on derived star-formation histories. We apply our prescription within the SED fitting code \texttt{Lightning} to a clean sample of 82, $z=0.21$--1.35 disk-dominated galaxies in the Great Observatories Origins Deep Survey North and South fields. To compare our inclination-dependent attenuation prescription with more traditional fitting prescriptions, we also fit the SEDs with the inclination-independent \citet{2000ApJ...533..682C} attenuation curve. From this comparison, we find that fits to a subset of 58, $z<0.7$ galaxies in our sample, utilizing the \citet{2000ApJ...533..682C} prescription, recover similar trends with inclination as the inclination-dependent fits for the far-UV-band attenuation and recent star-formation rates. However, we find a difference between prescriptions in the optical attenuation ($A_V$) that is strongly correlated with inclination ($p\textrm{-value}<10^{-11}$). For more face-on galaxies, with $i\lesssim50^{\circ}$, (edge-on, $i \approx 90^{\circ}$), the average derived $A_V$ is $0.31\pm0.11$ magnitudes lower ($0.56\pm0.16$ magnitudes higher) for the inclination-dependent model compared to traditional methods. Further, the ratio of stellar masses between prescriptions also has a significant ($p\textrm{-value}<10^{-2}$) trend with inclination. For $i=0^\circ$--$65^{\circ}$, stellar masses are systematically consistent between fits, with $\log_{10}(M_\star^{\rm inc}/M_\star^{\rm Calzetti}) =-0.05\pm0.03$ dex and scatter of 0.11 dex. However, for $i \approx 80^\circ$--$90^\circ$, derived stellar masses are lower for the \citet{2000ApJ...533..682C} fits by an average factor of $0.17\pm0.03$ dex and scatter of 0.13 dex. Therefore, these results suggest that SED fitting assuming the \citet{2000ApJ...533..682C} attenuation law potentially underestimates stellar masses in highly inclined disk-dominated galaxies.

\end{abstract}



\section{Introduction}

It is well understood that some fraction of the ultraviolet (UV) through near-infrared (NIR) light from stars is absorbed and reprocessed by dust into infrared (IR) and submillimeter emission within the interstellar media of galaxies \citep{1983A&A...128..212M,2003ARA&A..41..241D}. The portion of light that is reprocessed depends upon inherent properties, such as the distribution of dust grain size and shape, chemical composition, and the density of the dust \citep{2004ApJS..152..211Z,2007ApJ...657..810D}. Additionally, the portion of reprocessed starlight depends upon the geometric properties of the host galaxy, one of them being the orientation of the disk \citep[i.e., inclination;][]{2001ApJ...551..269G,2004A&A...419..821T,2011piim.book.....D,2013MNRAS.432.2061C}. For example, as the viewing angle of a galactic disk changes from face-on to edge-on (i.e., $i=0^\circ$ to $i=90^\circ$), the proportion of light that is processed along the line of sight increases due to an increasing column density of dust. This effect results in increased attenuation of highly inclined disk galaxies compared to low inclination galaxies \citep[e.g.,][]{1994AJ....107.2036G,2007MNRAS.379.1022D,2008ApJ...687..976U,2010MNRAS.404..792M,2011MNRAS.417.1760W,2016MNRAS.459.2054D,2017ApJ...851...90B,2018ApJ...859...11S}.

Accounting for the variation in attenuation due to inclination is crucial when determining the physical properties of galaxies. Both \citet{2008MNRAS.388.1708G} and \citet{2010ApJ...714L.113S} independently found that the inclination effects of dust can bias measurements of the galaxy $B$-band surface brightness to be $\approx$0.5$~\rm{mag~arcsec}^{-2}$ brighter for edge-on galaxies. Measurements of the half light radius have been shown to be increased by up to 110\% for edge-on galaxies compared to face-on galaxies \citep{2006A&A...456..941M,2018A&A...615A...7L}. UV magnitudes have been shown to be 1--2 magnitudes fainter for edge-on galaxies. This leads to underestimating the recent star-formation rates (SFR) by factors of 2.5--4 when using UV SFR calibrations \citep{2018MNRAS.480.3788W,2018ApJ...869..161W,2018A&A...616A.157L}. Conflicting results have been found for the effect of inclination on measurements of stellar mass. \citet{2009ApJ...691..394M} and \citet{2017MNRAS.468L..31D} report stellar mass to be almost independent at all inclinations, whereas \citet{2007MNRAS.379.1022D} and \citet{2018MNRAS.480.3788W} consider it inclination-independent from face-on to $\approx$70$^{\circ}$, above which masses can be underestimated by a factor of $\approx$2.

The same inclination-based attenuation applies when modeling the spectral energy distributions (SEDs) of galaxies (see \citealt{2013ARA&A..51..393C} for a review). Modeling SEDs allows for the derivation of the star formation histories (SFHs) of galaxies, from which the stellar mass and recent SFR are determined. In order to derive these properties from the observed SED, an attenuation curve is applied to stellar population synthesis models to construct attenuated model SEDs. These model SEDs are then fit to the observed SED to estimate the galaxy's SFH, and subsequently the stellar mass and recent SFR.

When determining the attenuated model SED, many studies utilize the \citet{2000ApJ...533..682C} attenuation law \citep[e.g.,][]{2015ApJ...801...97S,2018MNRAS.480.1973K,2019ApJS..243...22B} or the \citet{2001ApJ...548..296W} extinction curves for the Milky Way (MW), Large Magellanic Cloud, and Small Magellanic Cloud \citep[e.g.][]{2019ApJ...881...18R}. These curves are relatively rigid with the main flexibility in the free parameter used for normalization (i.e., $A_V$). A more flexible attenuation curve example is that from \citet{2009A&A...507.1793N}, which consists of a \citet{2000ApJ...533..682C} curve modified to include a UV bump and variable slope. Curves such as these are used to provide extra flexibility when fitting SEDs \citep[e.g.][]{2009A&A...507.1793N,2016A&A...591A...6B,2017ApJ...851...10E}, but they lack a direct physically motivated link to the inclination. High-spatial-resolution imaging surveys can provide constraints on the disk inclination and aid in accounting for the effects of inclination-based attenuation. However, these constraints would need to have a direct physically motivated link in the attenuation curve to properly be utilized.

In this paper, we utilize the inclination-dependent attenuation curves from \citet{2004A&A...419..821T} as updated by \citet{2011A&A...527A.109P} when fitting SEDs as to evaluate the effects of inclination on the derived SFHs. These physically motivated attenuation curves are based on radiative transfer calculations that use the commonly assumed dust composition of \citet{2007ApJ...663..866D} and geometries for the stellar and dust distributions that were shown to reproduce local observed galaxy SEDs \citep{2004A&A...419..821T,2011A&A...527A.109P}. The structure of the paper is as follows. In Section~\ref{sec:Data}, we describe the data and sample selection. In Section~\ref{sec:GalInc}, the method for estimating each galaxy's inclination is presented. In Section~\ref{sec:SEDModeling}, we describe our SED fitting procedure and the \citet{2004A&A...419..821T} inclination-dependent attenuation curve. In Section~\ref{sec:Results}, we present the results from the SED fittings. In Section~\ref{sec:Discussion}, we discuss the effects of inclination on the derived SFHs, specifically the recent SFR and stellar mass. Lastly, a summary is provided in Section~\ref{sec:Summary}.

For this study, we assume a \citet{2001MNRAS.322..231K} initial mass function with solar metallicity ($Z=Z_\odot$) and adopt a cosmology with $H_0=70~\rm{km~s}^{-1}~\rm{Mpc}^{-1}$, $\Omega_M=0.30$, and $\Omega_{\Lambda}=0.70$.

\section{Data and Sample Selection} \label{sec:Data}

To test the inclination-dependent attenuation prescription and study the resulting effects of inclination on the derived SFHs, we required a sample of galaxies that has high-quality uniform broadband data, spanning from the UV to far-infrared (FIR), and Hubble Space Telescope (HST) imaging data from which disk inclinations can be derived. The Great Observatories Origins Deep Survey (GOODS) North (N)and South (S) fields are excellent extragalactic survey fields for our study as they contain over 70,000 galaxies with deep \textit{HST} coverage and supplemental UV to FIR data \citep{2004ApJ...600L..93G}.

\subsection{Photometry} \label{sec:Photometry}

We utilized the UV to mid-infrared (MIR) photometry\footnote{Retrieved from the \textit{Rainbow} database: \url{http://rainbowx.fis.ucm.es/Rainbow_navigator_public/}} from \citet{2019ApJS..243...22B} and \citet{2013ApJS..207...24G} within the Cosmic Assembly Near-infrared Deep Extragalactic Legacy Survey (CANDELS) regions \citep{2011ApJS..197...35G,2011ApJS..197...36K} in the GOODS-N and GOODS-S fields, respectively. Both fields contain observations taken with HST Advanced Camera for Surveys (ACS) F435W F435W, F606W, F775W, F814W, and F850LP; HST Wide Field Camera 3 (WFC3) F105W, F125W, and F160W; and Spitzer Infrared Array Camera (IRAC) channels 1--4. The GOODS-N field also includes HST/WFC3 F140W, and the GOODS-S field includes HST/WFC3 F098M. The UV and NIR are supplemented by Kitt Peak National Observatory (KPNO) 4 m/Mosaic $U$, Large Binocular Telescope (LBT)/Large Binocular Camera (LBC) $U$, Subaru Multi-Object InfraRed Camera and Spectrograph (MOIRCS) $K_s$, and Canada France Hawaii Telescope (CFHT) Wide-field InfraRed Camera (WIRCam) $K_s$ ground-based observations for the GOODS-N; and Cerro Tololo Inter-American Observatory (CTIO) Blanco/Mosaic II $U$, Very Large Telescope (VLT)/ Visible Multi-Object Spectrograph (VIMOS) $U$, VLT Infrared Spectrometer And Array Camera (ISAAC) $K_s$, and VLT High Acuity Wide field K-band Imager (HAWK-I) $K_s$ ground-based observations for the GOODS-S. The methods from \citet{2019ApJS..243...22B} and \citet{2013ApJS..207...24G} for producing the photometry are the same and are briefly summarized below. The photometry and its uncertainty were extracted in all HST bands by running \texttt{SExtractor} \citep{1996A&AS..117..393B} in dual-image mode after identifying sources in the WFC3/F160W mosaic using a two-step “cold” plus “hot” strategy, as described in \citet{2013ApJS..206...10G} and \citet{2013ApJS..207...24G}. Source searching and photometry were performed after smoothing all other bands to the WFC3/F160W point-spread function (PSF). The lower resolution ground-based and Spitzer/IRAC photometry were determined using \texttt{TFIT} \citep{2007PASP..119.1325L} with the WFC3/F160W mosaic as the template image.

The FIR photometry used in our study was produced by \citet{2019ApJS..243...22B} for both the GOODS-N and GOODS-S fields and contains Spitzer Multiband Imaging Photometer (MIPS) 24 and 70~$\mu$m bands; Herschel Photodetector Array Camera and Spectrometer (PACS) 100 and 160~$\mu$m bands; and Herschel Spectral and Photometric Imaging Receiver (SPIRE) 250, 350, and 500~$\mu$m bands. To briefly summarize their methods, the FIR photometry associated with F160W sources consists of merged FIR photometric catalogs built from the data sets presented in \citet{2005ApJ...630...82P,2008ApJ...675..234P,2010A&A...518L..15P}, PACS Evolutionary Probe (PEP) + GOODS-Herschel \citep{2011A&A...532A..90L,2013A&A...553A.132M}, and Herschel Multi-tiered Extragalactic Survey \citep[HerMES;][]{2012MNRAS.424.1614O}. Due to the relatively low spatial resolution of the IR data, a cross-matching procedure was run from high (F160W) to low (SPIRE 500~$\mu$m) resolution bands as to obtain a one-to-one match for each F160W source. The most likely counterpart to a given IR source in the F160W image was chosen based on brightness and proximity to the IR source. A full description of the methods can be found in Appendix D of \citet{2019ApJS..243...22B}. We note that even though there should be minimal confusion of source identification for the PACS and SPIRE counterparts to the F160W sources, photometric issues could potentially arise due to nearby IR-bright sources. We discuss these issues and their potential effects on our final sample in Appendix~\ref{sec:IRAssess}.

Next, we corrected the photometry of each filter for Galactic extinction as estimated by the NASA Extragalactic Database extinction law calculator\footnote{\url{https://ned.ipac.caltech.edu/extinction_calculator}}, which uses the \citet{2011ApJ...737..103S} recalibration of the \citet{1998ApJ...500..525S} Cosmic Background Explorer (COBE) Diffuse Infrared Back- ground Experiment (DIRBE) and Infrared Astronomical Satellite (IRAS) Sky Survey Atlas (ISSA) dust maps. This recalibration assumes a \citet{1999PASP..111...63F} reddening law with $R_V = 3.1$. Our extinction values were determined for the center of each field, and we do not account for any variations across each of the GOODS fields, since extinction corrections for both fields are small and variation across the fields are minimal. These values, the corresponding filters used in each field, and the mean wavelength of the filters are listed in Table~\ref{table:SED}.

To include unaccounted for sources of uncertainty and systematic variations in the photometry, we added calibration uncertainties to the measured flux uncertainties that were derived by \texttt{SExtractor}, as is common when fitting SEDs \citep[e.g.,][]{2016A&A...591A...6B,2017ApJ...837..170L,2017ApJ...851...10E,2019ApJ...876....3L}. These calibration uncertainties are listed for each filter in Table~\ref{table:SED} as $\sigma_{C}$, which are the calibration uncertainties of 2--15\% as described by each instrument's user handbook. Further, we included 10\% model uncertainties for each band when fitting the SEDs to account for systematic effects in the models \citep{2016MNRAS.462.1415C,2019ApJS..240....3H}.

\subsection{Galaxy Sample Selection} \label{sec:SampleSelect}

\movetabledown=0.6in
\begin{deluxetable*}{lccc|lccc}
\tabletypesize{\footnotesize}
\tablecaption{Multiwavelength Coverage Used in SED Fitting \label{table:SED}}
\tablewidth{0pt}
\tablecolumns{12}
\tablehead{ \multicolumn{4}{c|}{\textbf{GOODS-North}} & \multicolumn{4}{c}{\textbf{GOODS-South}} \\\hline Instrument/Band & $\lambda_{\rm mean}$\tablenotemark{a} & $A_{\lambda}$\tablenotemark{b} & $\sigma_{C}$\tablenotemark{c} & Instrument/Band & $\lambda_{\rm mean}$\tablenotemark{a} & $A_{\lambda}$\tablenotemark{b} & $\sigma_{C}$\tablenotemark{c} \\ & ($\mu$m) & (mag) & & & ($\mu$m) & (mag) & }
\startdata
KPNO 4m/Mosaic $U$          & 0.3561  & 0.052  & 0.05  & Blanco/MOSAIC II $U$       & 0.3567  & 0.034  & 0.05  \\
LBT/LBC $U$                 & 0.3576  & 0.052  & 0.10  & VLT/VIMOS $U$              & 0.3709  & 0.033  & 0.05  \\
\textit{HST}/ACS F435W      & 0.4689  & 0.041  & 0.02  & \textit{HST} /ACS F435W    & 0.4689  & 0.027  & 0.02  \\
\textit{HST}/ACS F606W      & 0.5804  & 0.031  & 0.02  & \textit{HST} /ACS F606W    & 0.5804  & 0.020  & 0.02  \\
\textit{HST}/ACS F775W      &﻿0.7656  & 0.020  & 0.02  & \textit{HST} /ACS F775W    &﻿0.7656  & 0.013  & 0.02  \\
\textit{HST}/ACS F814W      &﻿0.7979  & 0.019  & 0.02  & \textit{HST} /ACS F814W    &﻿0.7979  & 0.012  & 0.02  \\
\textit{HST}/ACS F850LP     & 0.8990  & 0.015  & 0.02  & \textit{HST} /ACS F850LP   & 0.8990  & 0.010  & 0.02  \\
\textit{HST}/WFC3 F105W     & 1.0451  & 0.012  & 0.02  & \textit{HST} /WFC3 F098M   & 0.9829  & 0.008  & 0.02  \\
\textit{HST}/WFC3 F125W     & 1.2396  & 0.009  & 0.02  & \textit{HST} /WFC3 F105W   & 1.0451  & 0.008  & 0.02  \\
\textit{HST}/WFC3 F140W     & 1.3784  & 0.007  & 0.02  & \textit{HST} /WFC3 F125W   & 1.2396  & 0.006  & 0.02  \\
\textit{HST}/WFC3 F160W    & 1.5302 & 0.006 & 0.02 & \textit{HST} /WFC3 F160W  & 1.5302 & 0.004 & 0.02 \\
CFHT/WIRCam $K_s$          & 2.1413 & 0.004 & 0.05 & VLT/HAWK-I $K_s$          & 2.1403 & 0.002 & 0.05 \\
Subaru/MOIRCS $K_s$        & 2.1442 & 0.004 & 0.05 & VLT/ISAAC $K_s$           & 2.1541 & 0.002 & 0.05 \\
Spitzer/IRAC1             \tablenotemark{d} & 3.5314 & 0.002 & 0.05 & Spitzer/IRAC1            \tablenotemark{d} & 3.5314 & 0.001 & 0.05 \\
Spitzer/IRAC2             \tablenotemark{d} & 4.4690 & 0.000 & 0.05 & Spitzer/IRAC2            \tablenotemark{d} & 4.4690 & 0.000 & 0.05 \\
Spitzer/IRAC3             \tablenotemark{d} & 5.6820 & 0.000 & 0.05 & Spitzer/IRAC3            \tablenotemark{d} & 5.6820 & 0.000 & 0.05 \\
Spitzer/IRAC4             \tablenotemark{d} & 7.7546 & 0.000 & 0.05 & Spitzer/IRAC4            \tablenotemark{d} & 7.7546 & 0.000 & 0.05 \\
Spitzer/MIPS 24 $\mu$m    \tablenotemark{e} & 23.513 & 0.000 & 0.05 & Spitzer/MIPS 24 $\mu$m   \tablenotemark{e} & 23.513 & 0.000 & 0.05 \\
Spitzer/MIPS 70 $\mu$m    \tablenotemark{e} & 70.389 & 0.000 & 0.10 & Spitzer/MIPS 70 $\mu$m   \tablenotemark{e} & 70.389 & 0.000 & 0.10 \\
Herschel/PACS 100 $\mu$m  \tablenotemark{e} & 100.05 & 0.000 & 0.05 & Herschel/PACS 100 $\mu$m \tablenotemark{e} & 100.05 & 0.000 & 0.05 \\
Herschel/PACS 160 $\mu$m  \tablenotemark{e} & 159.31 & 0.000 & 0.05 & Herschel/PACS 160 $\mu$m \tablenotemark{e} & 159.31 & 0.000 & 0.05 \\
Herschel/SPIRE 250 $\mu$m \tablenotemark{e} & 247.21 & 0.000 & 0.15 & Herschel/SPIRE 250 $\mu$m\tablenotemark{e} & 247.21 & 0.000 & 0.15 \\
\enddata
\tablenotetext{a}{Mean wavelength of the filter calculated as $\lambda_{\textrm{mean}} = \frac{\int \lambda T(\lambda) d\lambda}{\int T(\lambda) d\lambda}$, where $T(\lambda)$ is the filter transmission function.}
\tablenotetext{b}{Galactic extinction for the center of the field.}
\tablenotetext{c}{Calibration uncertainties as given by the corresponding instrument user handbook.}
\tablenotetext{d}{Required band for the dust emission SED.}
\tablenotetext{e}{At least two of these bands are required for the dust emission SED, one of which must be $>100 \mu$m in the rest frame.}
\end{deluxetable*}

Since our goal is to present our inclination-dependent attenuation prescription and study the resulting effects of inclination on derived SFHs, we required a clean sample of disk-dominated galaxies, which our inclination-dependent analysis would apply. This sample was not required to be complete, but was limited to sources with high-quality data and unambiguous morphological types. Therefore, we initially selected, from the $\sim$70,000 galaxies within the GOODS fields, the 5459 galaxies with reliable spectroscopic redshifts \citep{2004ApJS..155..271S,2004AJ....127.3121W,2005A&A...437..883M,2006ApJ...653.1004R,2007A&A...465.1099R,2008ApJ...689..687B,2008A&A...478...83V,2009A&A...494..443P,2010A&A...512A..12B,2010ApJ...719..425F,2011AJ....141....1T,2012MNRAS.425.2116C,2015ApJS..218...15K}. Photometric redshifts are available for the galaxies that do not have spectroscopic redshifts \citep[e.g.,][]{2013ApJ...775...93D,2013ApJS..207...24G,2014ApJS..214...24S,2019ApJS..243...22B}. However, these photometric redshifts were derived from SED fittings and often have large uncertainties. Therefore, we do not include galaxies with photometric redshifts in our sample as the large uncertainties could have significant effects on our results. 

Inclination-dependent studies like ours can suffer from potential selection effects \citep{2016MNRAS.459.2054D}. We checked to see if requiring spectroscopic redshifts introduced any clear bias in our sample by preferentially selecting edge-on galaxies with elevated intrinsic luminosity distributions compared to face-on galaxies. Since spectroscopic redshift surveys are limited by the optical magnitude, often in the $r$-band, edge-on galaxies would need to be intrinsically more luminous compared to face-on galaxies to be above the magnitude limits, due to edge-on galaxies having higher optical/UV attenuation. This bias did not seem to be present in our final sample. For instance, the attenuated $r$-band absolute magnitudes of galaxy subsets in the final sample, binned by redshift, showed that nearly edge-on galaxies were fainter by 1--2 mag compared to face-on galaxies in the same redshift bin. This implies that the intrinsic luminosity distributions of the face-on and edge-on galaxies should be similar once attenuation had been removed, since edge-on galaxies would be more highly attenuated. We confirmed that the nearly edge-on ($1-\cos i>0.8$) and face-on ($1-\cos i<0.3$) intrinsic luminosity distributions were similar by performing a two-sided Kolmogorov-Smirnov (KS) test using the derived rest-frame $r$-band intrinsic luminosities and the inclinations from the SED fits, which results in a $p$-value $>0.3$ (see Section~\ref{sec:SEDIncEst}).

\begin{figure}[t!]
\centerline{
\includegraphics[width=8.75cm]{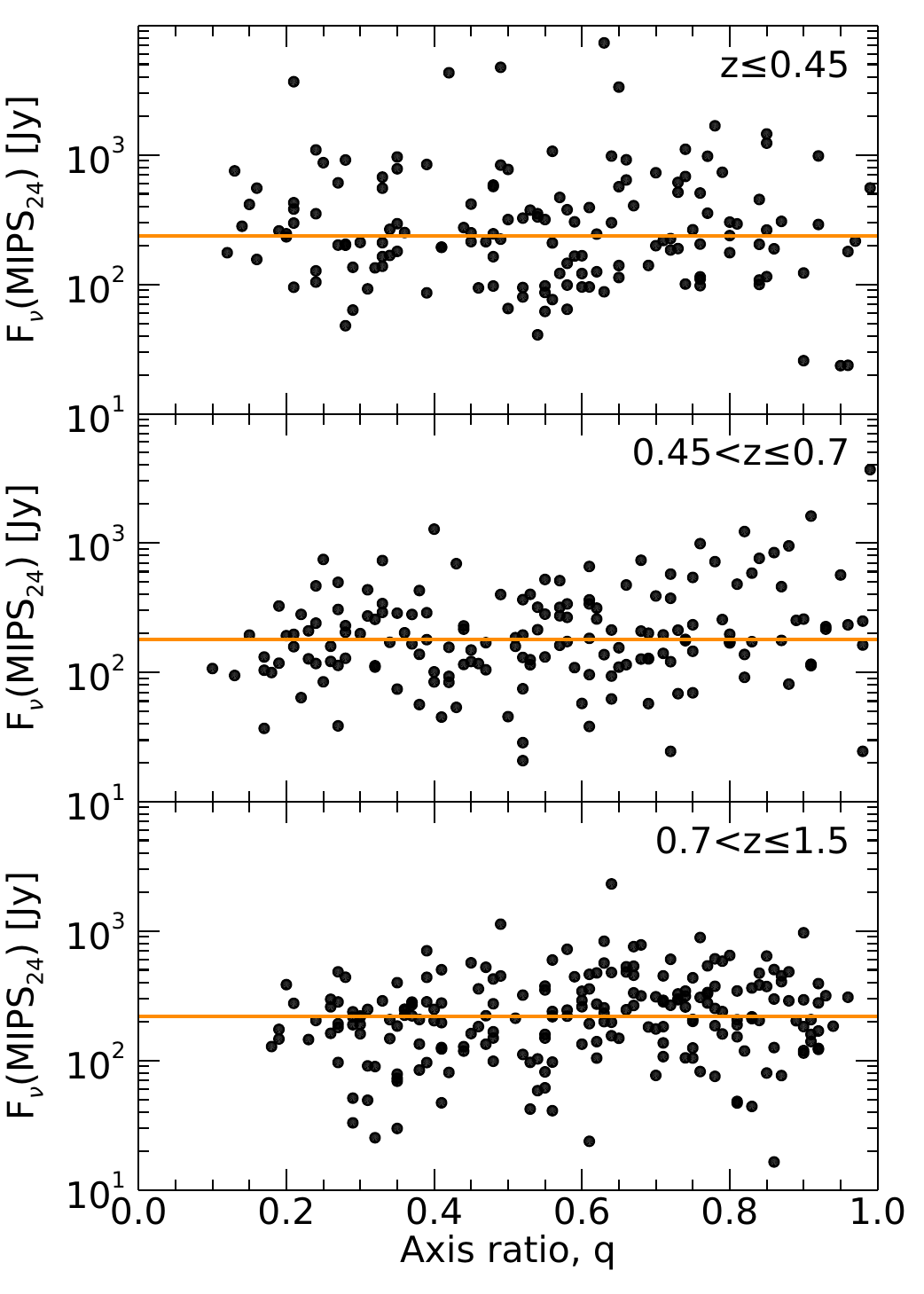}
}
\caption{
Spitzer/MIPS 24~$\mu$m fluxes as a function of axis ratio $q$ binned by redshift $z$. The orange lines in each panel represent the median 24~$\mu$m flux for that redshift bin. All three redshift bins can be seen to have no significant trends in the 24~$\mu$m fluxes vs. axis ratio, implying that the IR luminosity distributions are similar across inclination. This is expected as the 24~$\mu$m is practically attenuation free. So, we would expect no difference between edge-on and face-on galaxies.
}
\label{fig:mipsvqred}
\end{figure}

Next, we further limited our sample to galaxies that have at least six photometric measurements in the mid-to-far IR (3--1,000~$\mu$m) to better constrain the shape of the dust emission component of the SED, which we discuss in Section~\ref{sec:Dust}. We required that each galaxy has detections in all Spitzer/IRAC bands and permit the remaining two or more bands to be any combination of the Spitzer/MIPS, Herschel/PACS, or Herschel/SPIRE 250~$\mu$m data, one of which must be beyond the 100~$\mu$m rest frame to constrain the peak of the dust emission \citep{2007ApJ...663..866D,2013ARA&A..51..393C,2020MNRAS.498.4192F}. The fluxes for each band were required to have $F_\nu/\sigma_\nu>2$, where $\sigma_\nu$ includes the flux calibration uncertainty, which results in an original signal to noise ratio $>3$. This strict limitation led to the removal of 4918 galaxies from the 5459 galaxy sample, leaving 541 galaxies. 

To check if the IR selection requirement introduced any bias in our sample by preferentially selecting more IR luminous edge-on galaxies compared to face-on galaxies, we plotted the MIPS 24~$\mu$m fluxes as a function of the axis ratio in galaxy subsets, binned by redshift, as seen in Figure~\ref{fig:mipsvqred}. It can be seen that the MIPS 24~$\mu$m fluxes are similarly distributed at all axis ratios for each redshift bin. Thus, the lack of obvious differences in the 24 $\mu$m flux distributions indicates that the edge-on and face-on galaxies in our sample have similar IR luminosities.
\begin{figure*}[t!]
\centerline{
\includegraphics[width=18cm]{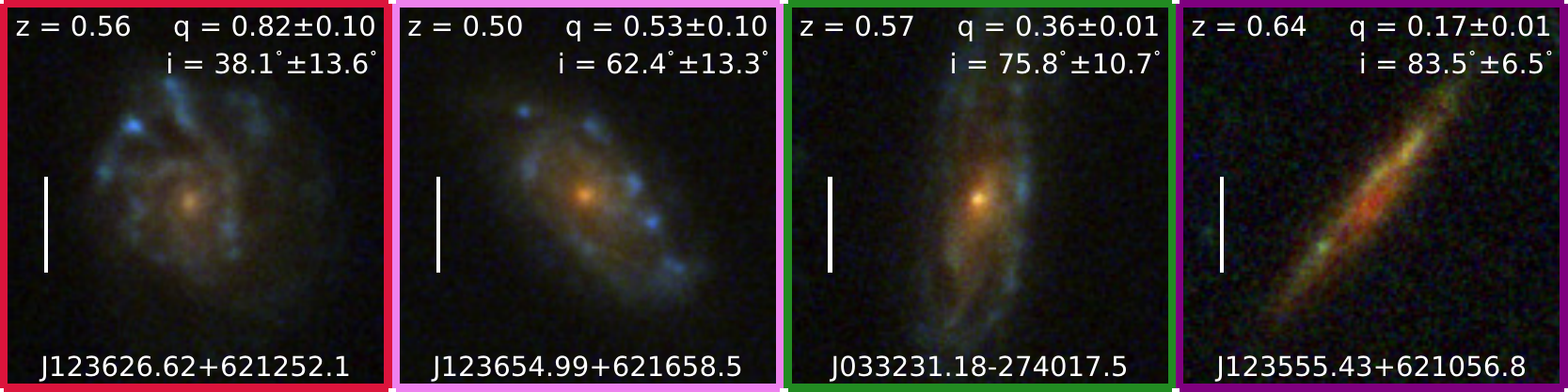}
}
\caption{
Composite HST/ACS F435W, F606W, and F850LP-band postage stamp images for galaxies within the final sample that were selected to span the full range of inclination $i$ and ordered by measured axis ratio $q$. Each stamp is centered on the source position, and a white $1^{\prime\prime}$ line is given for reference. NOTE: The outline color of each postage stamp is used to distinguish the respective galaxy in all subsequent figures that use these example galaxies.
}
\label{fig:postagestamp}
\end{figure*}

We further limited the sample to purely star-forming galaxies by identifying and removing sources that are flagged as active galactic nuclei (AGNs) from the Chandra X-ray catalogs for the GOODS-N \citep{2016ApJS..224...15X} and GOODS-S \citep{2017ApJS..228....2L} fields. Sources from the X-ray catalogs were matched to the CANDELS catalogs' sources with a matching radius of $1^{\prime\prime}$. We further attempted to limit the potential AGNs in our sample by removing obscured MIR-AGNs using the \citet{2012ApJ...748..142D} IRAC selection criteria and \citet{2013ApJ...763..123K} Spitzer/Herschel color-color criteria. A total of 114 potential AGNs were removed, leaving 427 galaxies.

Since our inclination-dependent attenuation prescription only applies to galaxies with disk morphologies, we limited our sample to only galaxies with clear disk morphologies. We selected disk galaxies using their S\'ersic index $n$ \citep{1963BAAA....6...41S}, where a galaxy is considered a disk galaxy if $n<2$. The S\'ersic indices for our galaxies were measured by \citet{2012ApJS..203...24V} using the \texttt{GALFIT} morphological code \citep{2002AJ....124..266P} on WFC3/F125W images in both the GOODS-N and GOODS-S fields. From these fits, 49 galaxies of the 427 remaining galaxies were not flagged as having a ``good fit'' (i.e., flag of 0) and were removed from the sample, leaving 378 galaxies.

Rather than using a S\'ersic index cutoff of $n<2$, we chose to further lower the cutoff to only include the 154 galaxies with $n<1.2$ out of the 378 remaining galaxies as to select disk-dominated (i.e., low $B/D$ ratio) galaxies. The choice of the cutoff value of $n<1.2$ is motivated by the work of \citet{2007ApJS..172..434S}, who showed that disk galaxies in the COSMOS field with purely exponential disks predominantly have $n<1.2$. The reason for selecting disk-dominated galaxies, rather than disk galaxies in general, is to reduce degeneracies within our SED fittings; we discuss this further in Section~\ref{sec:IncDepFits}. 

To confirm the selection of disk-dominated galaxies, we visually inspected the 154 galaxies that met the above criteria to confirm that there was no significant bulge and a clear disk was present. Since we limited our sample to strictly contain disk-dominated galaxies, any galaxy that could potentially be confused with an elliptical or irregular galaxy was removed from the sample. We also identified galaxies that appeared to have companions and may have been undergoing a merger, and removed these from our sample as well. In total, we chose to remove 72 galaxies from the sample that did not pass the visual inspection, and the final sample contains 82 galaxies spanning a redshift range of $z=0.21$--$1.35$. In Section~\ref{sec:Results}, we derive a mass range of $M_{\star}=10^{9.1}$--$10^{11.3} M_\odot$ and a SFR range of $\rm{SFR}=0.3$--$170\ M_\odot\ yr^{-1}$ for our sample and show that our galaxies are close to the redshift-dependent galaxy main sequence \citep[e.g.,][see Figure~\ref{fig:sfms}]{2015ApJ...801...80L}.

\begin{figure}[t!]
\centerline{
\includegraphics[width=8.75cm]{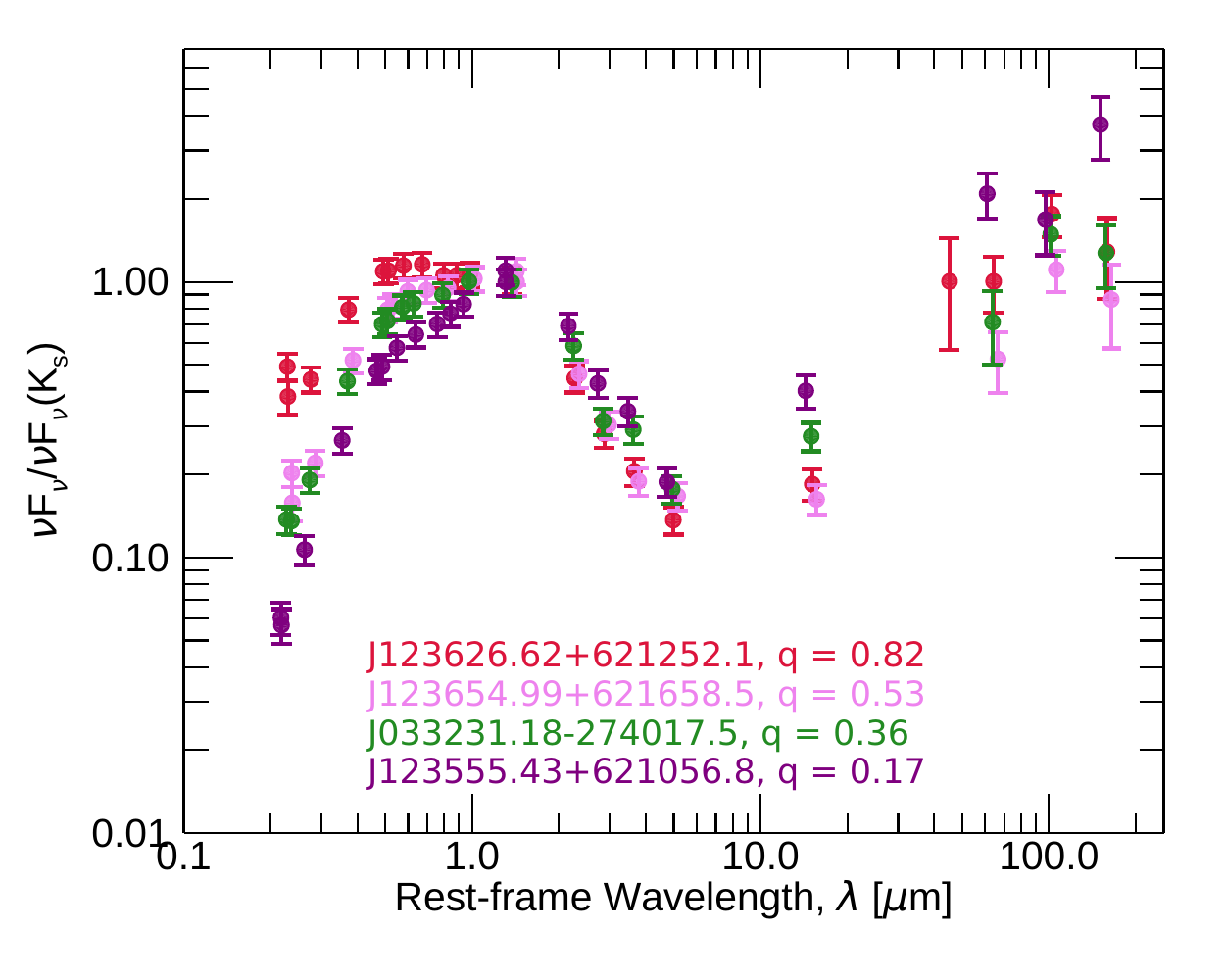}
}
\caption{
Broadband SEDs of the four example galaxies shown in Figure~\ref{fig:postagestamp} using the same color as the outline of the corresponding postage stamp. The SEDs are normalized to the Subaru/MOIRCS $K_s$ or VLT/ISAAC $K_s$ bands for the GOODS-N and GOODS-S galaxies, respectively. These SEDs show that as the axis ratio $q$ decreases that the UV-optical emission decreases, due to increasing attenuation.
}
\label{fig:obssed}
\end{figure}

Figure~\ref{fig:postagestamp} shows a set of the composite postage stamp images for galaxies that were selected to span the full range of inclination within the final sample. These galaxies will be used for illustrative purposes throughout the rest of the paper. The observed broadband SEDs for these sources are shown in Figure~\ref{fig:obssed} normalized to the Subaru/MOIRCS $K_s$ or VLT/ISAAC $K_s$ bands for the GOODS-N and GOODS-S galaxies, respectively. It can been seen that as the axis ratio $q$ decreases (i.e., inclination increases, see Equation \ref{eq:cosi}) that the UV-optical emission decreases, due to increased attenuation.

\section{Galaxy Inclinations} \label{sec:GalInc}

\begin{figure*}[t!]
\centerline{
\includegraphics[width=18cm]{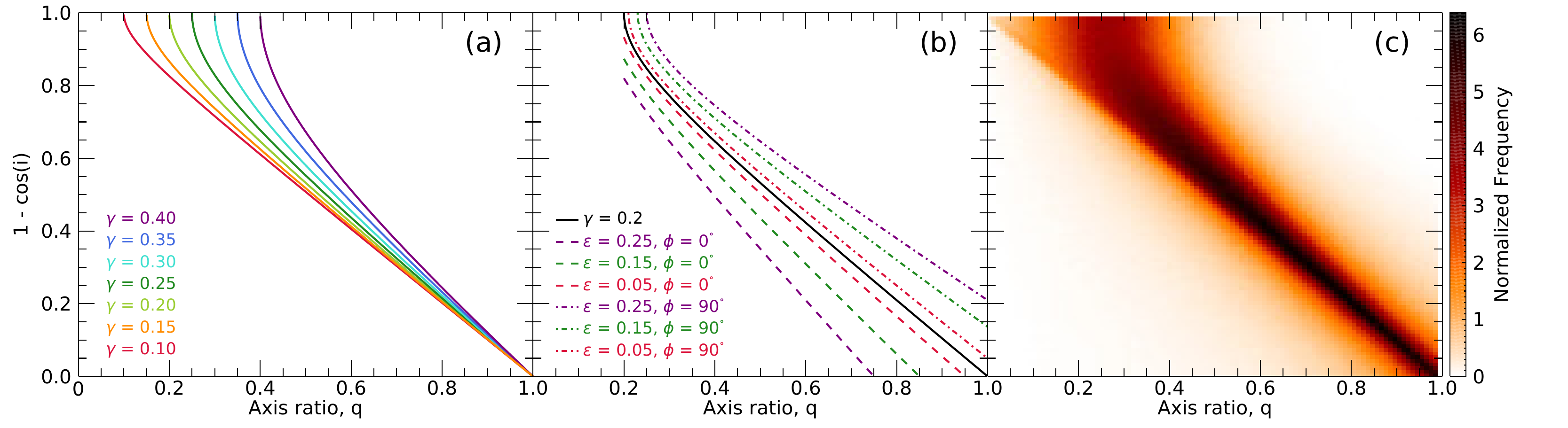}
}
\caption{
Each panel shows inclination ($1-\cos i$) as a function of axis ratio $q$. {(\textit{a})} The variation in possible inclinations for the same value of $q$ for different values of $\gamma$, a measure of the intrinsic thickness of the disk, within the observed range of disk galaxies represented as different colored lines. {(\textit{b})} The variation in possible inclinations for the same value of $q$ due to asymmetries, where the asymmetries causes $q$ to vary by a factor of $\sim\!1/(1-\epsilon\cos2\phi)$. The calculated inclinations for symmetric galaxies from Equation~\ref{eq:cosi} assuming $\gamma=0.2$ is represented as the solid black line for reference. {(\textit{c})} The two-dimensional distribution of inclination and $q$ from the Monte Carlo simulation.
}
\label{fig:qMC}
\end{figure*}

The inclination, $i$, of a disk galaxy is normally defined as the angle between the plane of the galactic disk and the plane of the sky. This means galaxies with $i=0^{\circ}$ and $i=90^{\circ}$ are considered face-on and edge-on, respectively. Inclination is difficult to measure directly and is normally derived from the axis ratio $q$ measured from an elliptical isophote or Se\'rsic profile. If galaxies were smooth, infinitely thin circular disks, then inclination could simply be determined by $\cos i=q$. However, galaxies have an intrinsic thickness ($\gamma$) when viewed edge-on, which is generally defined as the ratio between the scale height and scale length. Using the measured axis ratio and intrinsic thickness, inclination can be derived using the formula from \citet{1926ApJ....64..321H},
\begin{equation}
\cos^2i=\frac{q^2-\gamma^2}{1-\gamma^2},
\label{eq:cosi}
\end{equation}
where $q$ is the measured axis ratio, and $\gamma$ is the intrinsic thickness, which has been found from observations to mainly be within the range of $0.1<\gamma<0.4$ \citep[e.g.,][]{2008MNRAS.388.1321P,2008ApJ...687..976U,2013MNRAS.434.2153R}. For our study, we used the axis ratios measured from the fits for the Se\'rsic index by \citet{2012ApJS..203...24V} on WFC3/F125W images when determining the inclination.

Variation in $q$ with rest-frame wavelength has been observed in galaxies \citep[e.g.,][]{2002AJ....124.1328D}, which means that $q$ has been found to vary at different redshifts for the same observed photometric band. We checked this potential variation by comparing the WFC3/F125W and WFC3/F160W axis ratios from \citet{2012ApJS..203...24V} as a function of the redshift.  We found that any variation in $q$ at these redshifts was masked by the uncertainties on $q$, which agrees with the same analysis by \citet{2014ApJ...792L...6V}. Therefore, the WFC3/F125W axis ratios that we used are reliable for our entire sample's redshift range.

Blurring of a galaxy in its image by the PSF can also have a possible influence on the derived value of $q$. If the angular size of the minor axis is smaller than the angular size of the FWHM of the PSF, an artificial increase in the minor axis could occur, resulting in an overestimated value of $q$. All of the galaxies in our sample have minor axes that are larger than the PSF FWHM of the WFC3/F125W filter, such that blurring would not significantly influence our values of $q$. The minor (major) axis sizes have a range of 0.19$^{\prime \prime}$--1.53$^{\prime \prime}$ (0.84$^{\prime \prime}$--3.70$^{\prime \prime}$), with a median of 0.60$^{\prime \prime}$ (1.58$^{\prime \prime}$), which is larger than the 0.18$^{\prime \prime}$ PSF FWHM of the WFC3/F125W filter. Therefore, the following method used for determining an inclination from a measured axis ratio is applicable to our galaxies, and we note that the method should only be applied to galaxies that have minor axes larger than the PSF FWHM.

There are two important sources of uncertainty when calculating inclination using Equation~\ref{eq:cosi}. The first is that the value of $\gamma$ will vary among galaxies. However, a single value of $\gamma$ is normally applied when calculating inclinations for a large sample. By using a single value of $\gamma$ for a whole sample, galaxies can have large deviations between their calculated and true inclinations if their true $\gamma$ is different from the assumed value. This is a larger source of uncertainty in edge-on galaxies, where the measured axis ratio is small. This effect is shown in Figure~\ref{fig:qMC}{(a)}, where the colored lines represent different possible values of $\gamma$ within the observed range. The minimal effect on face-on galaxies is due to the intrinsic thickness of these galaxies not influencing the measured axis ratio as a result of the viewing angle. However, the difference in the intrinsic thickness of inclined galaxies can influence their measured axis ratio and lead to incorrect inclinations up to 23$^{\circ}$.

The second source of uncertainty comes from the fact that Equation~\ref{eq:cosi} assumes that galaxies are radially symmetric. However, it is apparent that galaxies are not radially symmetric, but instead have at least minor asymmetries due to clumpiness or spiral arms. It has been shown that asymmetries can cause the measured value of $q$ to vary from a radially symmetric value by a factor of $\sim\!1/(1-\epsilon\cos2\phi)$ , where $\epsilon$ is the intrinsic ellipticity (i.e., ellipticity of the disk due to asymmetries when viewed face-on) and $\phi$ is the azimuthal viewing angle relative to the intrinsic long axis of the disk \citep{2006ApJ...641..773R,2008ApJ...687..976U}. Changing the value of $\phi$ can be thought of as rotating a galaxy about the axis perpendicular to the plane of the disk such that the intrinsic ellipticity causes the measured axis ratio to vary depending on whether the major or minor axis of the intrinsic ellipticity is aligned to the viewing angle. If $\epsilon$ and $\phi$ are known, Equation~\ref{eq:cosi} could be updated by replacing $q$ with $q/(1-\epsilon\cos2\phi)$ to recover the correct inclination. However, $\epsilon$ and $\phi$ are rarely known for deep-field galaxies and are often ignored when determining the inclination. An example of how this source of uncertainty affects the inclination can be seen in Figure~\ref{fig:qMC}{(b)} for the case of $\gamma=0.2$.

To determine inclinations for the galaxies in our sample in a way that incorporates these sources of uncertainty, we ran a Monte Carlo simulation to determine each galaxy's inclination probability density function (PDF). As stated above, if galaxies were infinitely thin circular disks, then inclination could simply be determined by $\cos i=q$. If they were randomly oriented, we would expect a uniform distribution with respect to $\cos i$ (see below) and therefore $q$. However, as shown above, $i$ is dependent upon $q$ as well as $\gamma$, $\epsilon$, and $\phi$. This leads to $q$ no longer being a uniform distribution, but rather being a function of the distributions of $\cos i$, $\gamma$, $\epsilon$, and $\phi$ given by
\begin{equation}
q=\left(\sqrt{\cos^2i\:(1-\gamma^2)+\gamma^2}\right)(1-\epsilon\cos2\phi).
\label{eq:MonteCarlo}
\end{equation}
Therefore, the goal of our Monte Carlo simulation is to determine the unknown distribution of $q$ using Equation~\ref{eq:MonteCarlo} from the known distributions of $\cos i$, $\gamma$, $\epsilon$, and $\phi$; from which a distribution of $i$ can be determined for a given value and uncertainty of $q$. 

For the distribution of inclination, $\cos i$ would be uniformly distributed if galaxies were randomly oriented. When observing a galaxy from a random direction, each solid angle element surrounding the galaxy from which to observe it would be equally likely. Comparatively, observing a galaxy at a given inclination could be thought of as viewing it from a solid-angle band (i.e., each inclination is a line of latitude on the surrounding celestial sphere). This band will cover larger areas at $i\approx90^{\circ}$ (i.e., equator) compared to $i\approx0^{\circ}$ (i.e., the poles). This larger area leads to more external galaxies viewing the galaxy at $i\approx90^{\circ}$ compared to $i\approx0^{\circ}$. In other words, there are more lines of sight for a nearly edge-on view than for a nearly face-on view of a galaxy. This leads to the probability of observing a galaxy being distributed by a sine function. Via the probability integral transform, this means $1-\cos i$ is uniformly distributed, and therefore $\cos i$ has a uniform distribution as well.

For $\phi$, we assumed a uniform distribution between its possible values of 0 and $2\pi$. As for $\epsilon$ and $\gamma$, we used the PDFs for these random variables given in Figure~11 of \citet{2013MNRAS.434.2153R}, who derived these distributions from 92,923 spiral galaxies with $r$-band data from the Sloan Digital Sky Survey (SDSS) Data Release~8 \citep[DR8;][]{2011ApJS..195...26A}, which had morphologies based on the Galaxy Zoo project \citep{2011MNRAS.410..166L}. The galaxies used in their SDSS DR8 sample were in the redshift range of $z=0$--$0.15$ with a median of $z\approx0.1$, while our sample galaxies' redshifts are $z=0.21$--$1.35$ with a median of $z\approx0.56$. In the rest frames, the $r$-band used in their study and the WFC3/F125W band used in our study are $\approx$0.56~$\mu$m and $\approx$0.79~$\mu$m, respectfully. These rest-frame bands are comparable, and therefore, the error introduced by using these PDFs, which are derived from a different photometric band than our data, is assumed to be negligible. We also tested two additional distributions of $\gamma$ and $\epsilon$ provided in Figure~11 of \citet{2013MNRAS.434.2153R}, which have smaller values of $\gamma$, and found negligible differences in the inclination distributions derived from the Monte Carlo simulations. However, we do note that the distributions of $\gamma$ may be skewed to higher values due to PSF blurring effects from the limited angular resolution of SDSS, especially when compared to the intrinsic thickness of nearby, highly resolved edge-on galaxies.

Further, the inclination-dependent \citet{2004A&A...419..821T} attenuation curves described in Section \ref{sec:AttCurve} assume $\gamma\approx0.08$ and $\epsilon=0$ at the rest-frame wavelength of $\approx$0.56~$\mu$m. This leads to an internal inconsistency with our model by using distributions of $\epsilon$ and $\gamma$ from \citet{2013MNRAS.434.2153R} rather than these fixed values. However, assuming a fixed value for these variables only decreases the uncertainty on the derived inclinations.

We ran the Monte Carlo simulation for $10^7$ trials to thoroughly sample the distribution. Each trial consisted of a draw from the distributions of $\cos i$, $\gamma$, $\epsilon$, and $\phi$, which resulted in a value of $q$ from Equation~\ref{eq:MonteCarlo}. We discarded $<6\%$ of the $10^7$ trials due to them resulting in $q>1$, which can occur when the simulated galaxy is nearly face-on ($\cos i \approx 1$) and $\cos2\phi < 0$. The resulting two-dimensional distribution of inclination and $q$ can be seen in Figure~\ref{fig:qMC}{(c)}. Having this two-dimensional distribution, we needed to determine each galaxy's inclination PDF from it in a way that incorporated how the uncertainty of the measured value of $q$ is distributed. This was done by generating an additional $10^6$ values of $q$ drawn from a Gaussian distribution whose mean and standard deviation were the measured value of $q$ and its uncertainty from \citet{2012ApJS..203...24V}. After removing any of the additional $10^6$ values of $q$ that exceeded the possible values of $q$, we matched them to their closest $q$ values from the Monte Carlo simulation and recorded the corresponding inclination values. Therefore, each galaxy's inclination PDF consisted of these $\sim$10$^6$ corresponding inclination values from the matched values of $q$.

\begin{figure}[t!]
\centerline{
\includegraphics[width=8.75cm]{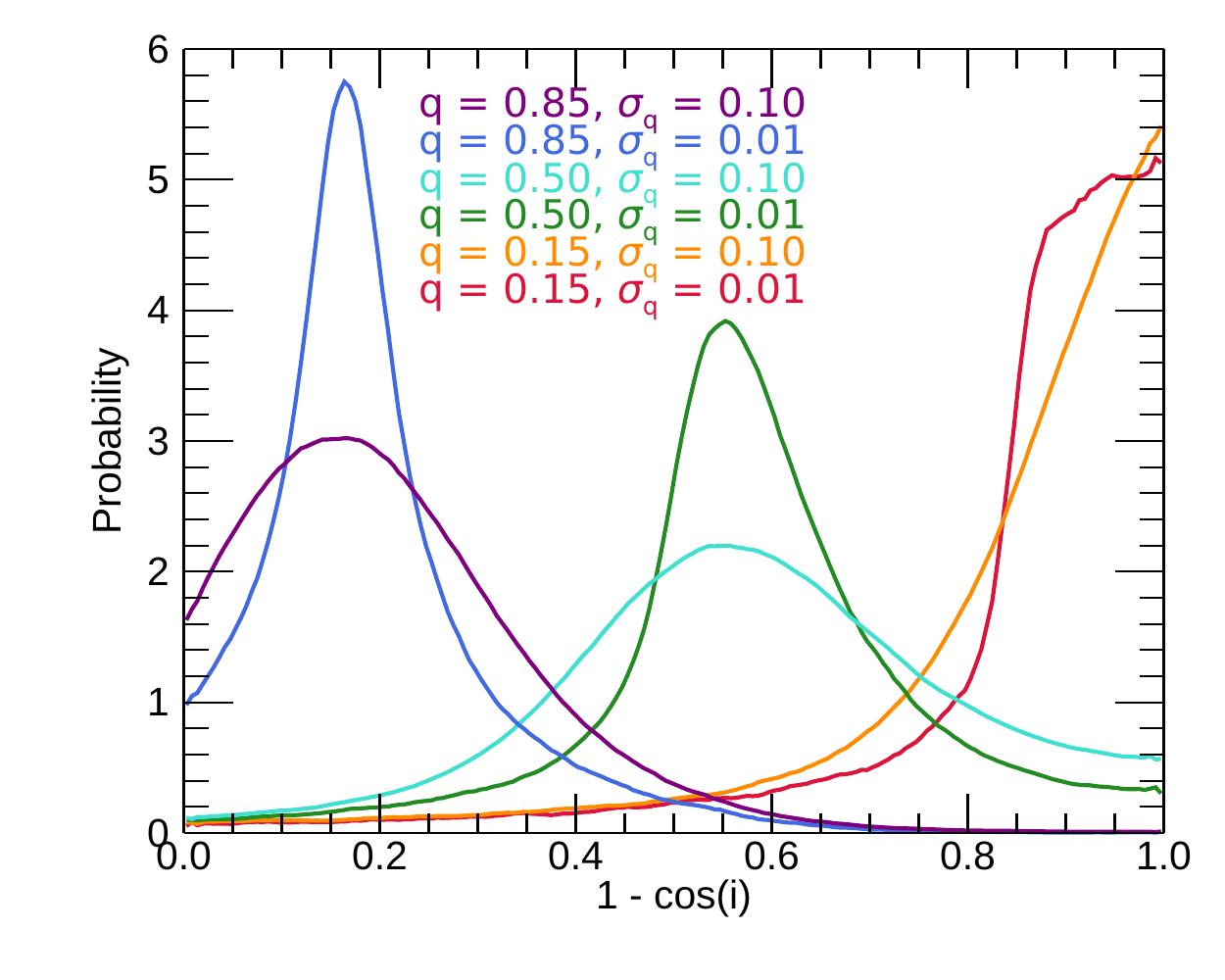}
}
\caption{
Probability distributions of inclination ($1-\cos i$) from the Monte Carlo simulation for different example distributions of $q$. The distributions consist of $q$ values of 0.85, 0.5, and 0.15 with $\sigma_q$ values of 0.1 and 0.01. As $\sigma_q$ increases for a fixed $q$, the width of the inclination distribution increases, since an increase in the uncertainty in $q$ expectedly increases the uncertainty in inclination.
}
\label{fig:IncPDF}
\end{figure}

Figure~\ref{fig:IncPDF} shows example inclination PDFs for $q=0.85$, 0.5, and 0.15 with standard deviations of $\sigma_q=0.1$ and 0.01. From these examples, it can be seen that as $\sigma_q$ increases for a fixed $q$, the width of the inclination distribution increases. This is expected, since as the uncertainty in $q$ increases, so should the uncertainty in inclination.

\begin{figure}[t!]
\centerline{
\includegraphics[width=8.75cm]{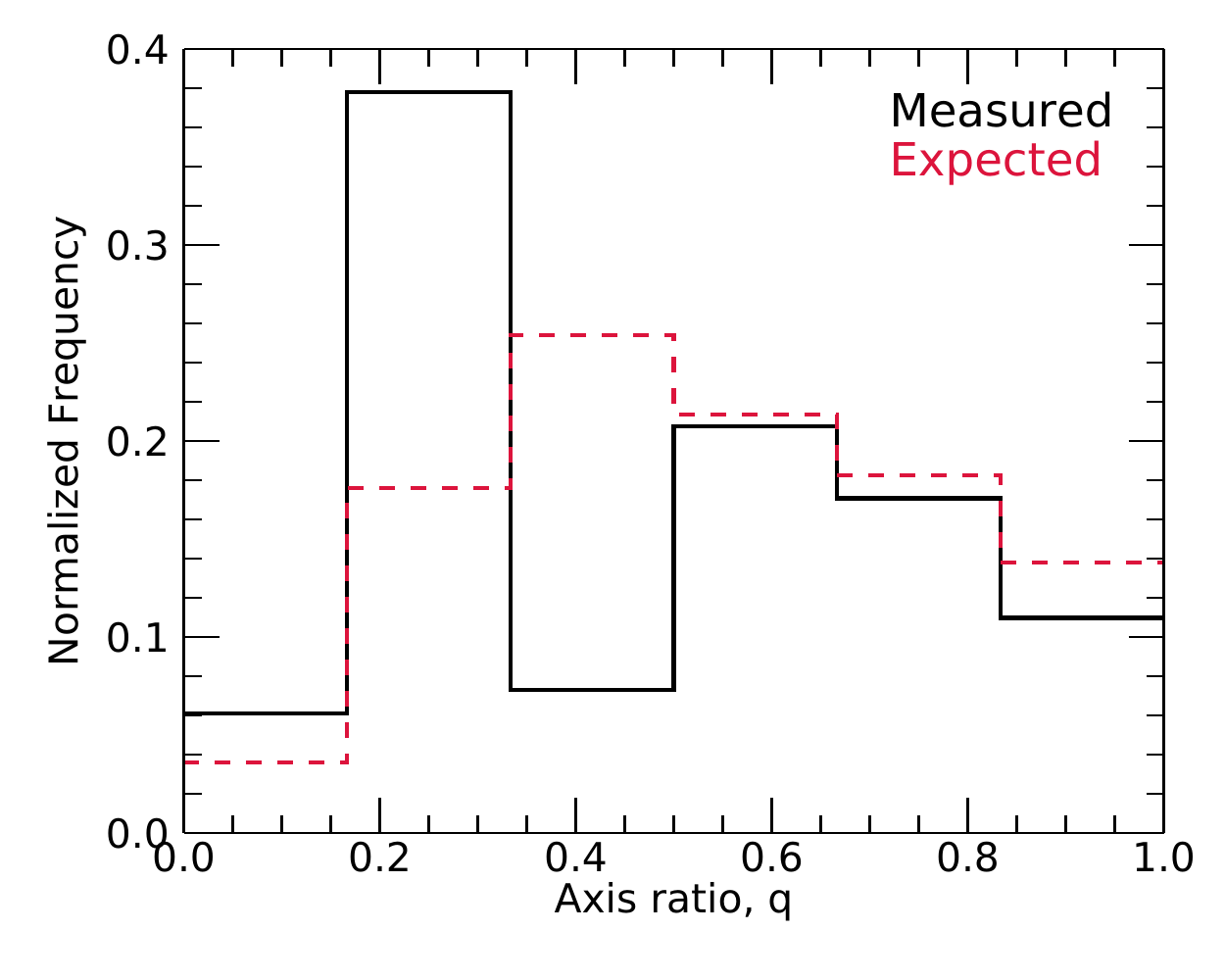}
}
\caption{
Distribution of axis ratio $q$. The solid black line shows the distribution of our galaxy sample using the measured values from \citet{2012ApJS..203...24V}, and the dashed red line shows the expected distribution from the Monte Carlo simulation if our sample comprised randomly oriented disk galaxies. The discrepancy between the two distributions is due to various effects in our sample selection. For example, the visual inspection likely leads to more edge-on galaxies, which are easier to visually distinguish as disk galaxies compared to moderately inclined galaxies that were more easily confused for elliptical galaxies.
}
\label{fig:qdist}
\end{figure}

Marginalizing the two-dimensional distribution of $q$ and $i$ for $q$ from the Monte Carlo simulation gives the expected distribution if galaxies are randomly oriented (i.e. uniform distribution in $\cos i$), which is shown as a dashed red line in Figure~\ref{fig:qdist}. The distribution of the measured axis ratios from \citet{2012ApJS..203...24V} for the galaxies in our sample is shown as the black line. The two distributions are statistically distinct (a two-sided KS test gives a $p$-value $<10^{-5}$), with our sample showing a deficit of moderately inclined galaxies as well as an excess of edge-on galaxies. However, this is expected, since we did not require a complete sample. For example, during the visual inspection, edge-on galaxies were more likely to be admitted into the sample as they are easier to visually distinguish as disk galaxies compared to moderately inclined galaxies, which were more easily confused for elliptical galaxies.

Finally, we quantified the consequences of not incorporating variation in $\gamma$ and assuming radial symmetry when determining inclination. We compared the median, 16th, and 84th percentiles of our sample's inclination PDFs to inclinations and uncertainties of our sample calculated using Equation~\ref{eq:cosi} assuming radial symmetry and the commonly used fixed values of $\gamma=0.15$ and 0.2 \citep[e.g.,][]{2009ApJ...691..394M,2010ApJ...714L.113S,2013MNRAS.432.2061C,2018ApJ...869..161W,2018A&A...615A...7L} as well as $\gamma=0.08$ assumed by \cite{2004A&A...419..821T} in deriving the inclination-dependent attenuation curves. These calculated values for the fixed values of $\gamma$ are in excellent agreement with the median PDF inclinations for galaxies with $i\gtrsim30^{\circ}$. For $i\lesssim30^{\circ}$, inclinations are slightly lower (2$^{\circ}$--15$^{\circ}$) for the calculated values due to not including asymmetries. Comparing the uncertainties, the calculated values uncertainties are underestimated by an average factor of $\approx$7.9 for $\gamma=0.15$, $\approx$7.5 for $\gamma=0.2$, and $\approx$8.8 for $\gamma=0.08$ compared to the PDF uncertainties. Therefore, if the variation in $\gamma$ is ignored and radial symmetry is assumed, the inclination can be properly recovered from Equation~\ref{eq:cosi} if $i\lesssim30^{\circ}$, but the uncertainty will be underestimated.

\section{SED Modeling} \label{sec:SEDModeling}

\subsection{SED Fitting Procedure} \label{sec:SEDFitPro}

To fit the SEDs of our galaxies, we used the SED fitting code \texttt{Lightning}  \footnote{Version 2.0 \url{https://github.com/rafaeleufrasio/lightning}} \citep{2017ApJ...851...10E}. \texttt{Lightning} is an SED fitting procedure that models the FUV to NIR stellar emission with \texttt{P\'EGASE}  population synthesis models \citep{1997A&A...326..950F}. The modeled stellar emission includes attenuation that can be restricted to be in energy balance with the integrated NIR to FIR (5--1000$\mu$m) dust emission. For this paper, we have updated \texttt{Lightning} to include a module that models this NIR to FIR dust emission with the dust models from \citet{2007ApJ...657..810D} (see Section~\ref{sec:Dust}).

The SFH model consists of five time steps at  0--10~Myr, 10--100~Myr, 0.1--1~Gyr, 1--5~Gyr, and 5--13.6~Gyr with each period having a constant SFR. The final age bin upper bound for a given galaxy was fixed to the age of the universe at that galaxy's redshift. If the age of the universe for a galaxy was less than 5~Gyr, then the fifth age bin was omitted, and the fourth age bin upper bound was fixed at to the age of the universe at that galaxy's redshift.  We list the possible and adopted ranges for the SFR of each bin and the assumed priors used when fitting the SEDs in Table~\ref{table:LightParam}.

\begin{deluxetable*}{lccr}
\tablecaption{Adjustable Parameters and Ranges within Lightning \label{table:LightParam}}
\tablecolumns{4}
\tablehead{ \\ Parameter & Possible Range & Range in This Work & Prior Distribution \\                     &   (Min, Max)   &     (Min, Max)     &   }
\startdata
\cutinhead{Star Formation History Bins [$M_\odot \ yr^{-1}$]}
$\psi_1$ (0-10 Myr) & ($0, \infty$) & ($0, \infty$) & Flat \\
$\psi_2$ (10-100 Myr) & ($0, \infty$) & ($0, \infty$) & Flat \\
$\psi_3$ (0.1-1 Gyr) & ($0, \infty$) & ($0, \infty$) & Flat \\
$\psi_4$ (1-5 Gyr)\tablenotemark{a} & ($0, \infty$) & ($0, \infty$) & Flat \\
$\psi_5$ (5-13.6 Gyr)\tablenotemark{a} & ($0, \infty$) & ($0, \infty$) & Flat \\
\cutinhead{\citet{2007ApJ...657..810D} Dust Emission Model}
$\alpha$            & ($-10$, 4)              & (2, 2)                         & Fixed \\
$U_{\rm min}$       & (0.1, 25)               & (0.7, 25)                      & Flat  \\
$U_{\rm max}$       & ($10^3$, $3\times10^5$) & ($3\times10^5$, $3\times10^5$) & Fixed \\
$\gamma_{\rm dust}$ & (0, 1)                  & (0, 1)                         & Flat  \\
$q_{\rm PAH}$       & (0.0047, 0.0458)        & (0.0047, 0.0458)               & Flat  \\
\cutinhead{\citet{2000ApJ...533..682C} Attenuation Law} 
$\tau_V^{\rm diff}$\tablenotemark{b} & (0, 3) & (0, 3) & Flat \\
\cutinhead{Inclination-dependent Attenuation Curves}
$\tau_B^{\rm f}$                 & (0, 8)        & (0, 8)                      & Flat  \\
$r^{0,\rm old}$\tablenotemark{c} & (0, 1)        & \ \ (0, 1)\tablenotemark{d} & Fixed\tablenotemark{d} \\
$B/D$                            & ($0, \infty$) & (0, 0)                      & Fixed \\
$F$                              & (0, 0.61)     & (0, 0.61)                   & Flat  \\
$\cos i$                         & (0, 1)        & (0, 1)                      & Flat/Image-based distribution\tablenotemark{e} \\
\enddata
\tablenotetext{a}{The age ranges of the oldest two age bins depend upon the redshift.}
\tablenotetext{b}{Proportional to $A_V$ ($\tau_V^{\rm diff}=0.4\ln10\,A_V$).}
\tablenotetext{c}{$r^{0,\rm old}$ is a binary parameter with 0 designating ``young' star formation history bins and 1 designating ``old' star formation history bins. The star formation history bins that contain ages $\lesssim$ 500~Myr are required to be considered ``young' (see Section~\ref{sec:AttCurve}). }
\tablenotetext{d}{For this work, we define the ``young' star formation history bins as those with look-back times $<1$~Gyr (i.e., $\psi_1$, $\psi_2$, and $\psi_3$) and the ``old' bins as those with look-back times $>1$~Gyr (i.e., $\psi_4$ and $\psi_5$).}
\tablenotetext{e}{The SEDs were fit twice with the inclination-dependent model. Once with the inclination prior as a flat distribution, and again with the prior as the image-based inclination distributions derived in Section \ref{sec:GalInc} (see Section \ref{sec:SEDIncEst}).}
\end{deluxetable*}

The time steps of the SFH model can be arbitrarily chosen in \texttt{Lightning}. However, our time steps were chosen such that the first step, 0--10~Myr, models the stellar population that is able to emit enough hydrogen-ionizing photons to produce noticeable hydrogen recombination lines. The second time step of 10--100~Myr was chosen to model the stellar population that emits the majority of the UV emission when combined with the first step. The combination of the first two time steps provides the average SFR of the past 100~Myr, which is a timescale commonly used by SFR calibrations \citep[e.g.,][]{1998ARA&A..36..189K,2007ApJ...666..870C,2011ApJ...741..124H}. The final three steps were chosen to have comparable bolometric luminosities to that of the second time step for the case of a constant SFR (see \citealt{2017ApJ...851...10E} for details).

Due to the relatively large number of free parameters used in this work (all of which are listed in Table~\ref{table:LightParam}), we added a module to \texttt{Lightning} that uses Markov Chain Monte Carlo (MCMC) analysis via the Metropolis-Hastings algorithm \citep{Metropolis1953,Hastings1970} to fit each SED and derive posterior probability densities of the SFH time steps, attenuation parameters, and dust model parameters. Due to the complex nature of our models, manually selecting an optimal covariance matrix for the sampled proposal multivariate normal distribution was challenging. Therefore, we also implemented a vanishing adaptive MCMC algorithm (see Algorithm 4 from \citealt{Andrieu2008}), which adaptively determines the optimal covariance matrix. It does this by modifying the covariance matrix with each step in the chain until an optimal acceptance ratio \citep{Gelman1996} is reached. This modification of the covariance matrix with previous steps is not a true Markov chain, due to the present being affected by the past. However, the vanishing part of the algorithm causes the amount of modification to the covariance matrix to decrease with each step in the chain. Therefore, with a long enough chain, the modification to the covariance matrix will cease, and the resulting ending segment of the chain will be a true Markov chain. In Section~\ref{sec:Results}, we further discuss the use of the MCMC procedure in estimating the parameter distributions.

The MCMC algorithm was added and utilized over the matrix inversion algorithm in the previous version of \texttt{Lightning} (v1.0), since the matrix inversion algorithm required a grid for the dust attenuation and emission parameters. Due to the increase in parameters from the dust emission model (see Section~\ref{sec:Dust}) and inclination-dependent attenuation (see Section~\ref{sec:AttCurve}), this method was no longer feasible due to very long computational times, whereas the MCMC algorithm run time is less sensitive to an increase in the number of parameters. For example, using the dust emission model and the inclination-dependent attenuation both with all parameters free, the MCMC algorithm with $10^5$ iterations takes approximately the same amount of time as the inversion method with a coarse grid of six points per parameter.

\subsection{Dust Emission Model} \label{sec:Dust}

Our goal for modeling the dust emission component of the SEDs in this paper is to retrieve the total infrared luminosities. To accomplish this, we use the \citet{2007ApJ...657..810D} dust model, which utilizes a mixture of carbonaceous and silicate grains, whose grain size distributions were made to be compatible with the extinction in the MW \citep{2001ApJ...548..296W}. The model parameterizes the dust mass exposed to the radiation field intensity $U$, which ranges from $U_{\rm min}$ to $U_{\rm max}$, as a superposition of a delta function at $U_{\rm min}$ and a power law of slope $\alpha$ between $U_{\rm min}$ and $U_{\rm max}$. This is given by Equation 23 in \citet{2007ApJ...657..810D},
\begin{align}
  \frac{dM_{\rm dust}}{dU} = & (1-\gamma_{\rm dust})M_{\rm dust} \delta(U-U_{\rm min})  \notag\\
    & +\gamma_{\rm dust} M_{\rm dust} \frac{(\alpha-1)}{(U_{\rm min}^{1-\alpha}-U_{\rm max}^{1-\alpha})}U^{-\alpha}, \ \ \alpha \neq 1,
\end{align}
where $M_{\rm dust}$ is the total dust mass, $\gamma_{\rm dust}$ is the fraction of dust mass exposed to the power-law radiation field, and $\delta$ is the Dirac $\delta$-function. There is one other relevant parameter in the model, $q_{\rm PAH}$, which is the polycyclic aromatic hydrocarbon (PAH) index. The PAH index is defined to be the fraction of the total grain mass corresponding to PAHs containing less than 1000 carbon atoms.

Excluding the normalization parameter $M_{\rm dust}$, there are five free parameters within the dust model: $\alpha$, $U_{\rm min}$, $U_{\rm max}$, $\gamma_{\rm dust}$, and $q_{\rm PAH}$. Of these parameters, three most strongly control the shape of the model IR SED: $q_{\rm PAH}$, $\gamma_{\rm dust}$, and $U_{\rm min}$ \citep{2007ApJ...663..866D,2017ApJ...837..170L}. As for $U_{\rm max}$ and $\alpha$, \citet{2007ApJ...663..866D} found that dust model fits are not very sensitive to precise values of these two parameters and that the IR SEDs of galaxies in the Spitzer Infrared Nearby Galaxies Survey \citep{2003PASP..115..928K} were well reproduced by $U_{\rm max}=~10^6$ and $\alpha=2$. Therefore, we adopt the fixed values of $U_{\rm max}=3\times10^5$ and $\alpha=2$ when fitting the SEDs as described in Section \ref{sec:Results}. We note that \citet{2007ApJ...663..866D} used $U_{\rm max}=10^6$ rather than $U_{\rm max}=3\times10^5$. However, our current set of dust models has a maximum $U_{\rm max}$ of $3\times10^5$. Therefore, we used this value instead and expect minimal difference in fittings, since $U_{\rm max}$ is insensitive to precise values.\footnote{\texttt{Lightning} computes the dust emission model using the publicly available $\delta$-functions of $U$, from which the power-law component can be calculated for any given $\alpha$. The largest available $\delta$-function of $U$ is $U=3\times10^5$. Therefore, rather than extrapolating to $U=10^6$, we limit $U$ to the largest available value.} The possible and adopted ranges for the dust emission parameters and the assumed priors used when fitting the SEDs can be seen in Table~\ref{table:LightParam}. We note that $M_{\rm dust}$ is not a free parameter in our models, rather the normalization of the dust emission is dependent upon the total attenuation via energy balance (see Section~\ref{sec:EnCon}).

\subsection{Inclination-dependent Attenuation Curves} \label{sec:AttCurve}

The original two FUV to NIR attenuation modules in \texttt{Lightning} were the original \citet{2000ApJ...533..682C} attenuation law as well as its modified version by \citet{2009A&A...507.1793N}, which includes a bump and a variable UV slope. To evaluate the effects of inclination on the derived SFHs, we required an inclination-dependent attenuation model. Therefore, we added another attenuation module that utilizes the inclination-dependent attenuation curves from \citet{2004A&A...419..821T}, as updated by \citet{2011A&A...527A.109P}.

To create the inclination-dependent attenuation curves, \citet{2004A&A...419..821T} used the ray-tracing radiative transfer code of \citet{1987ApJ...317..637K} to determine the attenuation of the stellar emission from disk galaxies at different inclinations. They used geometries for the stellar and dust distributions that were shown to reproduce observed galaxies' UV to submillimeter SEDs. The model geometry consists of an exponential disk of old stars with associated diffuse dust (disk), a dustless old de Vaucouleurs stellar bulge (bulge), a thin exponential disk of young stars with associated diffuse dust that represents the stars and dust within spiral arms (thin disk), and a clumpy dust component that represents the dense molecular clouds within the star-forming regions of the thin disk (clumpy component). The dust model originally used by \citet{2004A&A...419..821T} was the graphite and silicate dust model of \citet{1993ApJ...402..441L}. However, the dust model was updated by \citet{2011A&A...527A.109P} to the dust model of \citet{2001ApJ...548..296W} and \citet{2007ApJ...657..810D}, which includes PAH molecules in addition to the graphite and silicate particles.

To determine the attenuation from the diffuse dust, \citet{2004A&A...419..821T} superposed the diffuse dust from each disk and derived the attenuation as seen through the combined dust disks for each geometric component (disk, thin disk, and bulge) at various combinations of inclinations, central face-on optical depths in the $B$-band (the optical depth of the galaxy in the $B$-band as seen through the center of the galaxy if it were face-on), $\tau_B^f$, and wavelengths. They then fit the resulting attenuation curves as a function of inclination (i.e., $\Delta m$ vs. $1-\cos i$) for each component, wavelength, and $\tau_B^f$ with fifth order polynomials, whose coefficients were made publicly available\footnote{\url{http://cdsarc.u-strasbg.fr/viz-bin/qcat?J/A+A/527/A109}}. The wavelength range spanned 0.0912 to 2.2~$\mu$m, and the sampled values of $\tau_B^f$ were 0.1, 0.3, 0.5, 1.0, 2.0, 4.0, and 8.0, which span the range of optically thin to thick.

The attenuation due to the clumpy component in the thin disk was determined analytically rather than with radiative transfer calculations. This was calculated by assuming there was some probability that light from stars would be absorbed by the star's parent molecular cloud. The calculation was represented as a clumpiness factor $F$, which is defined as the total fraction of UV light that is locally absorbed by the parent cloud. This clumpiness factor is independent of the galaxy inclination, due to it being a local, rather than a global, galactic phenomenon.

The inclination-dependent attenuation for a whole galaxy is calculated by combining each geometric and clumpy component attenuation at a given wavelength and is given by
\begin{multline}
\label{eq:TuffTotAtt}
\Delta m_{\lambda} = -2.5 \log \Bigg(r^{0,\rm disk} 10^\frac{\Delta m_{\lambda}^{\rm disk}(i,\tau_B^f)}{-2.5} \\
+ (1-r^{0,\rm disk}-r^{0,\rm bulge})(1-F f_{\lambda}) 10^\frac{\Delta m_{\lambda}^{\rm tdisk}(i,\tau_B^f)}{-2.5} \\
+ r^{0,\rm bulge} 10^\frac{\Delta m_{\lambda}^{\rm bulge}(i,\tau_B^f)}{-2.5}  \Bigg),
\end{multline}
where $\Delta m_{\lambda}$ is the composite attenuation at a given wavelength $\lambda$; $r^{0,\rm disk}$ and $r^{0,\rm bulge}$ are the fractions of the intrinsic flux densities from the disk and bulge components, respectively, relative to the total intrinsic flux density of the galaxy; $\Delta m_{\lambda}^{\rm disk}(i,\tau_B^f)$, $\Delta m_{\lambda}^{\rm tdisk}(i,\tau_B^f)$, and $\Delta m_{\lambda}^{\rm bulge}(i,\tau_B^f)$ are the attenuation from the diffuse dust given by the fifth order polynomials that are a function of inclination for a tabulated $\tau_B^f$ at the given wavelength for the disk, thin disk, and bulge, respectively; $F$ is the clumpiness factor; and $f_{\lambda}$ is a tabulated function of wavelength which gives $F$ its wavelength dependence. Further, the two parameters, $r^{0,\rm disk}$ and $r^{0,\rm bulge}$, can be redefined by two, more intuitive parameters, the fraction of intrinsic flux density from the old stellar components compared to the total intrinsic flux density $r^{0,\rm old}$ and the $B/D$ ratio, which are given by
\begin{align}  
    r^{0,\rm old}&= r^{0,\rm disk}+r^{0,\rm bulge}, \label{eq:rold} \\ 
    B/D&=r^{0,\rm bulge}/r^{0,\rm disk}. \label{eq:bd}
\end{align}
Therefore, since $\Delta m_{\lambda}^{\rm disk}(i,\tau_B^f)$, $\Delta m_{\lambda}^{\rm tdisk}(i,\tau_B^f)$, and $\Delta m_{\lambda}^{\rm bulge}(i,\tau_B^f)$ are dependent upon the inclination and $\tau_B^f$, the five wavelength-independent free parameters of our attenuation curves are $i$, $\tau_B^f$, $r^{0,\rm old}$, $B/D$, and $F$.

We note that $B/D$ defined here is the ratio of the intrinsic luminosity of the old stellar bulge to the old stellar disk. Yet, measured values of $B/D$ for galaxies are the observed luminosity ratio of the bulge to the disk. Therefore, since we do not necessarily expect the attenuation in the observed band for each of these components to be the same, the measured $B/D$ could vary from the expected input $B/D$. Further, the observed emission from the disk will include emission from the young stellar thin disk as well. This inclusion of the thin disk can bias the measured $B/D$ to smaller values than the input $B/D$ parameter should be. However, both of these potential biases can be mitigated if the $B/D$ for a galaxy is measured from a rest-frame NIR band (i.e., $J$, $H$, or $K$), where attenuation and the contribution from the young stellar population should both be minimal.

In the original equation given by \citet{2004A&A...419..821T}, $r^{0,\rm disk}$ and $r^{0,\rm bulge}$ are observable rather than intrinsic properties (i.e., fraction of observed flux densities from the disk or bulge components compared to the total observed flux density) and wavelength dependent, with this wavelength dependence being used to vary the weight of each component at a given wavelength. However, by having $r^{0,\rm disk}$ and $r^{0,\rm bulge}$ as intrinsic properties and combining them into $r^{0,\rm old}$ and $B/D$, we can take advantage of our nonparametric SFH to effectively eliminate the need for a wavelength dependence and $r^{0,\rm old}$ as a free parameter. This is done by setting $r^{0,\rm old}=0$ for all SFH age bins that are considered to be young populations and $r^{0,\rm old}=1$ for those that are considered to be old populations. With these criteria, we assume that the young stellar population in the SFH is contained within the thin disk, and the older populations are within the disk and bulge.  If $r^{0,\rm old}$ was allowed to be a free parameter, it would require a wavelength dependence to properly account for how the young and old populations contribute to the total emission at each wavelength. Since this would be computationally expensive, we set $r^{0,\rm old}$ as a fixed binary parameter in the attenuation curves, leaving four free parameters  $i$, $\tau_B^f$, $B/D$, and $F$. 

We note that when designating SFH age bins as young and old populations for the binary parameter $r^{0, \rm old}$, any age bin that contains ages $\lesssim$ 500~Myr should be considered part of the young population. This is required due to the assumption by \citet{2004A&A...419..821T} that only the young population in the thin disk emits in the UV, and therefore, the old stellar population attenuation curve components ($\Delta m_{\lambda}^{\rm disk}(i,\tau_B^f)$ and $\Delta m_{\lambda}^{\rm bulge}(i,\tau_B^f)$) are zero for UV wavelengths ($\lambda \lesssim 4430~\rm{\AA}$). Since stellar models in \texttt{Lightning} with ages $\lesssim$ 500~Myr can significantly contribute to the unattenuated UV emission, we require any age bin containing ages $<$ 500~Myr to be considered part of the young population as to have this significant UV emission attenuated. Stellar models with ages $>$ 500~Myr have 2--3 orders of magnitude lower unattenuated UV emission than those with ages $\lesssim$ 500~Myr at the same SFR and do not significantly contribute to the total UV emission even when unattenuated. However, we strongly emphasize that this will only be the case when there is a prevalent young population, such as in our galaxy sample. If a galaxy has a highly dominant older population, then the UV emission from this population could dominate the observed UV, and the assumption by \citet{2004A&A...419..821T} that the old stellar population has no UV attenuation would no longer hold.\footnote{It is possible to extrapolate $\Delta m_{\lambda}^{\rm disk}(i,\tau_B^f)$ and $\Delta m_{\lambda}^{\rm bulge}(i,\tau_B^f)$ into the UV, as shown in \citet{2013MNRAS.432.2061C}. However, implementing an extrapolation is beyond the scope of this paper, but it will be pursued in future work.}

To compute the total attenuation from Equation~\ref{eq:TuffTotAtt}, we first calculated the attenuation from each geometric component $\Delta m_{\lambda}^{\rm disk}(i,\tau_B^f)$, $\Delta m_{\lambda}^{\rm tdisk}(i,\tau_B^f)$, and $\Delta m_{\lambda}^{\rm bulge}(i,\tau_B^f)$ using the tabulated polynomial coefficients from \citet{2011A&A...527A.109P} for each tabulated wavelength and $\tau_B^f$, for an input inclination. To the tabulated wavelengths and values of $\tau_B^f$, we added the wavelength of 5.0~$\mu$m and $\tau_B^f=0$ for later interpolation smoothness. For these new tabulated values, we set the attenuation of each geometric component to zero. This is because at $\tau_B^f=0$ there should be no attenuation from the diffuse dust, and we adopted 5.0~$\mu$m to be the cutoff wavelength above which there will be no attenuation, because it matched the longest tabulated wavelength of $f_{\lambda}$ in Table E.4 of \citet{2011A&A...527A.109P}.

\begin{figure*}[t!]
\centerline{
\includegraphics[width=18cm]{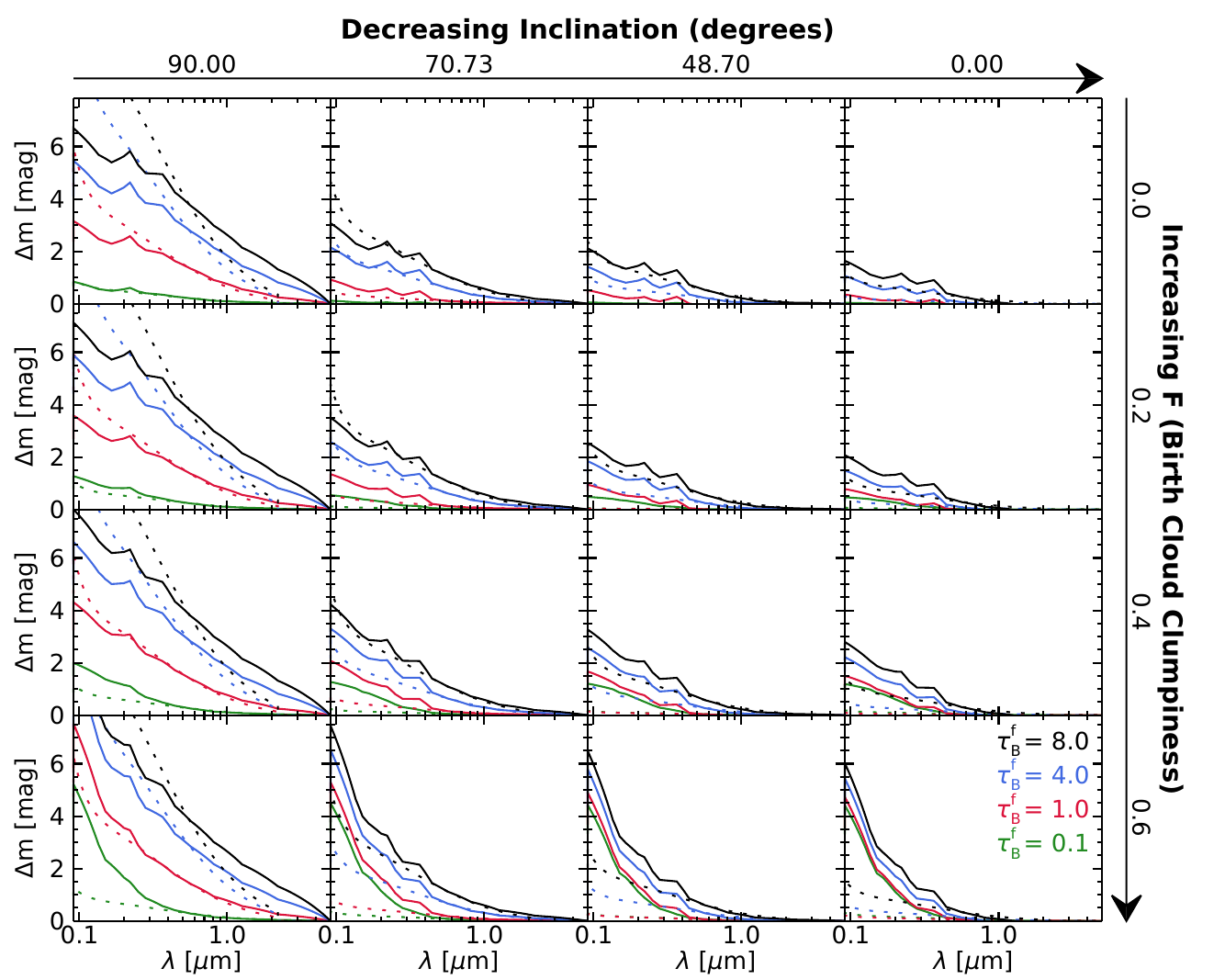}
}
\caption{
Each panel shows the attenuation as a function of the wavelength at four values of $\tau_B^f$, the central face-on optical depth in the $B$-band (\textit{solid} colored lines). Each curve has $r^{0,\rm old} = 0$ and $B/D = 0$, which are the fraction of intrinsic flux densities from the old components compared to the total intrinsic flux density of the galaxy and the $B/D$ ratio. The panels from left to right show how the attenuation is affected by decreasing inclination, with the inclination values being equally spaced in $\cos i$ space. The panels from top to bottom show how the birth cloud clumpiness $F$ causes the UV attenuation to become steeper. The \textit{dotted} lines show the \citet{2000ApJ...533..682C} attenuation curve normalized to the same $A_V$ (0.55~$\mu$m) as the corresponding \textit{solid} colored line for comparison.
}
\label{fig:tuffsattenuation}
\end{figure*}

Next, we calculated $\Delta m_{\lambda}$ from Equation~\ref{eq:TuffTotAtt} with the precomputed values of $\Delta m_{\lambda}^{\rm disk}(i,\tau_B^f)$, $\Delta m_{\lambda}^{\rm tdisk}(i,\tau_B^f)$, and $\Delta m_{\lambda}^{\rm bulge}(i,\tau_B^f)$ for an input $r^{0,\rm old}$ and $B/D$ (converted to $r^{0,\rm disk}$ and $r^{0,\rm bulge}$ by rearranging Equations \ref{eq:rold} and \ref{eq:bd}) and $F$ along with the tabulated values of $f_{\lambda}$. This resulted in $\Delta m_{\lambda}$ as an array of values corresponding to the tabulated values of wavelength and $\tau_B^f$. Finally, we interpolated this array for an input $\tau_B^f$ and input wavelengths to determine the total attenuation at the input wavelengths. To assure that there is no erroneous extrapolation beyond our tabulated wavelength range, we set the total attenuation to zero for wavelengths not within the range of $0.0912~\mu{\rm m} \le \lambda \le 5.0~\mu{\rm m}$. The possible and adopted ranges for each attenuation parameter and the assumed priors used when fitting the SEDs are listed in Table \ref{table:LightParam}.

We note that \citet{2004A&A...419..821T} recommends interpolating $\Delta m_{\lambda}^{\rm disk}(i,\tau_B^f)$, $\Delta m_{\lambda}^{\rm tdisk}(i,\tau_B^f)$, and $\Delta m_{\lambda}^{\rm bulge}(i,\tau_B^f)$ for $\tau_B^f$ and the wavelength, and interpolating $f_{\lambda}$ for wavelength before using Equation~\ref{eq:TuffTotAtt}. However, we found that our method is faster computationally by a factor of 2 without any significant differences in the $\Delta m_{\lambda}$ values. Therefore, the inclination-dependent attenuation module in \texttt{Lightning} interpolates after using Equation~\ref{eq:TuffTotAtt}.

\begin{figure*}[t!]
\centerline{
\includegraphics[width=18cm]{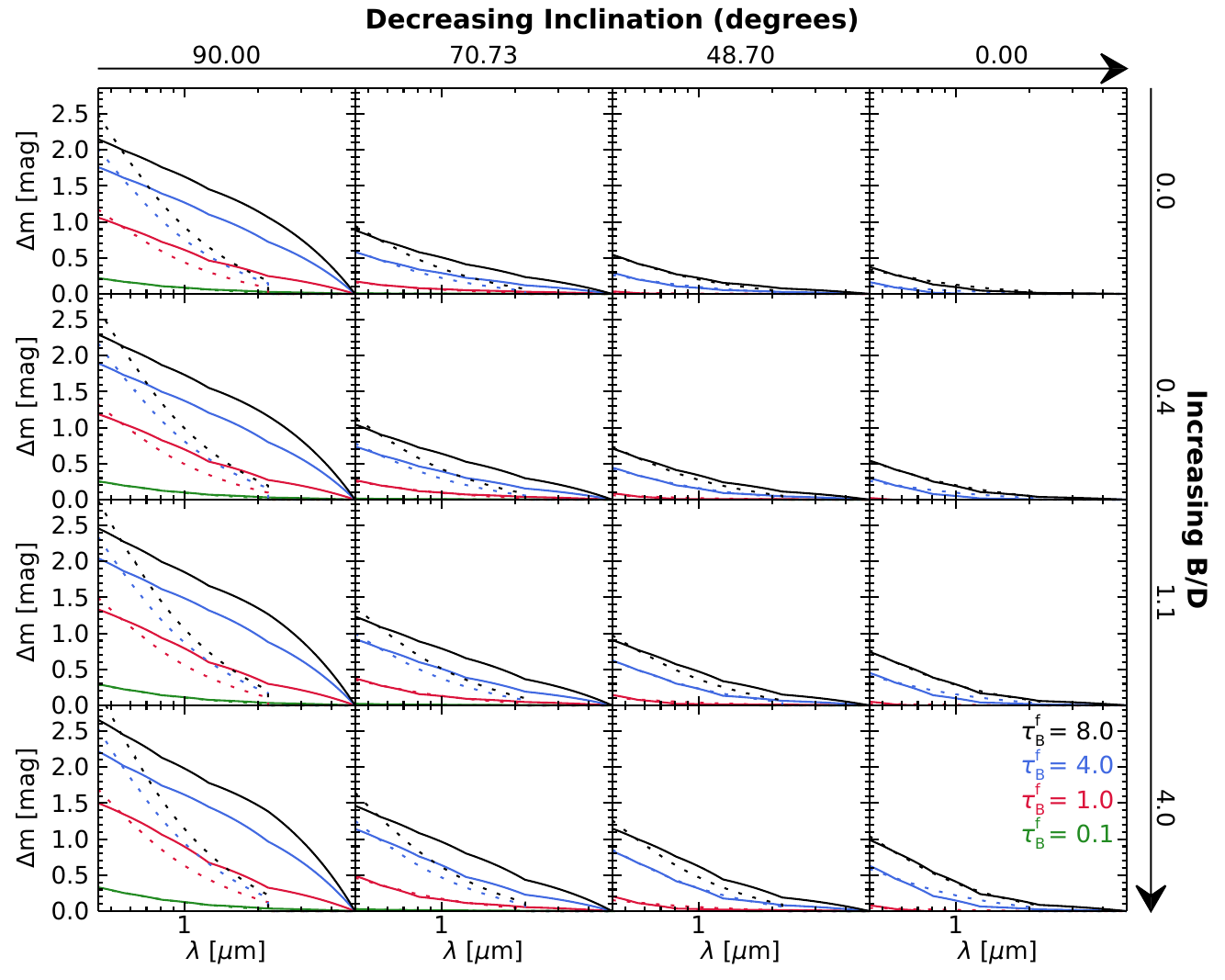}
}
\caption{
Each panel shows the attenuation as a function of the wavelength at four values of $\tau_B^f$, the central face-on optical depth in the $B$-band (\textit{solid} colored lines). Each curve has $r^{0,\rm old} = 1$ and $F = 0$, which are the fraction of intrinsic flux densities from the old components compared to the total intrinsic flux density of the galaxy and the clumpiness factor. The panels from left to right show how the attenuation is affected by decreasing inclination, with the inclination values being equally spaced in $\cos i$ space. The panels from top to bottom show how attenuation is affected by increasing the $B/D$ ratio, with the $B/D$ values being equally spaced in the bulge-to-total $B/T$ space ($B/D=B/T/(1-B/T)$). The truncation of the attenuation curves at 0.443~$\mu$m is due to the assumption by \citet{2004A&A...419..821T} that the old stellar population does not emit light at wavelengths shorter than 0.443~$\mu$m and therefore does not have attenuation. The \textit{dotted} lines show the \citet{2000ApJ...533..682C} attenuation curve normalized to the same $A_V$ (0.55~$\mu$m) as the corresponding \textit{solid} colored line for comparison.
}
\label{fig:tuffsattenuationrold}
\end{figure*}

Examples of the young population (i.e., $r^{0,\rm old} = 0$ and $B/D = 0$) attenuation curves for the span of $\tau_B^f$, $F$, and inclination are shown as the solid curves in Figure~\ref{fig:tuffsattenuation}. The increase in $\tau_B^f$ with the other parameters fixed gives the expected result of steeper attenuation curves. As inclination increases to edge-on, the attenuation curves again become steeper. However, inclination also has the more influential effect, compared to $\tau_B^f$, of causing attenuation at longer wavelengths. For face-on galaxies, wavelengths beyond 1.0~$\mu$m are negligibly attenuated, but edge-on galaxies can be significantly attenuated out to the attenuation curve limit of 5.0~$\mu$m. The clumpiness component $F$ can be seen to steepen the attenuation curves in the UV, while leaving the optical attenuation relatively unchanged.

The dotted curves in Figure~\ref{fig:tuffsattenuation} show the original \citet{2000ApJ...533..682C} attenuation law for comparison. The normalization of each curve is set to the same $A_V$ (0.55~$\mu$m) as the corresponding solid colored line in each panel. The \citet{2000ApJ...533..682C} attenuation law has only one free parameter, the diffuse $V$-band optical depth $\tau_V^{\rm diff}$, which is proportional to $A_V$. The possible and adopted range for $\tau_V^{\rm diff}$ and its assumed prior used when fitting the SEDs are listed in Table \ref{table:LightParam}. We note that $\tau_V^{\rm diff}$ differs in definition from $\tau_B^f$, beyond being in different optical bands. The parameter $\tau_V^{\rm diff}$is defined as the average optical depth over all solid angles, whereas $\tau_B^f$ is defined as the optical depth through the center of the galaxy, the location with the maximum dust surface density, when viewed face-on. In Figure \ref{fig:tuffsattenuation}, comparisons between the solid and dotted lines of matching color show the rigidity of the \citet{2000ApJ...533..682C} curve compared to the inclination-dependent curves. Also from the comparison, it can be seen that the \citet{2000ApJ...533..682C} curve rarely aligns with the inclination-dependent attenuation curves, especially in cases of edge-on inclinations and high birth cloud clumpiness.

Figure~\ref{fig:tuffsattenuationrold} shows example attenuation curves of the old population (i.e., $r^{0,\rm old}=1$ and $F=0$) for the span of $\tau_B^f$, $B/D$, and inclination as the solid curves. The attenuation curves are truncated at wavelengths shortward of 0.443~$\mu$m due to the assumption by \citet{2004A&A...419..821T} that the old stellar population does not provide substantial emission at wavelengths shorter than 0.443~$\mu$m and therefore does not have attenuation. As with the young population curves, an increase in $\tau_B^f$ with the other parameters fixed gives steeper attenuation curves. Increasing inclination to edge-on, the attenuation curves again steepen and attenuation also occurs at longer wavelengths. Increasing the $B/D$ with the other parameters fixed results in steeper attenuation curves similar to increasing $\tau_B^f$. Comparing to the dotted curves, which show the original \citet{2000ApJ...533..682C} attenuation law normalized to the same $A_V$ as the corresponding solid colored line in each panel, it can be seen that the \citet{2000ApJ...533..682C} attenuation law has a very similar shape as the low inclination curves for all $B/D$ values at optical wavelengths. However, as with the young population curves, the curves diverge as inclination approaches edge-on.

\subsection{Energy Balance/Conservation} \label{sec:EnCon}

Energy balance in SED fitting is the assumption that the power absorbed by attenuating dust is equal to the radiative power of the dust emission \citep[i.e., the UV through NIR attenuated light is reemitted in the IR and submillimeter; e.g.,][]{2008MNRAS.388.1595D,2017ApJ...837..170L,2019A&A...622A.103B,2019A&A...632A..79B}. However, energy balance is not true energy conservation, due to it considering the line-of-sight intensity as representative of the isotropic power rather than the total $4\pi$ steradian anisotropic integrated power. As stated above, the attenuation in disk galaxies is not equivalent at all viewing angles, but depends on the inclination. Therefore, to apply more realistic energy conservation, an inclination-dependent attenuation curve can be used to account for the line-of-sight variation of the attenuated emission and aid in determining the total bolometric power.

When applying any of the attenuation modules to the stellar emission, \texttt{Lightning} can require energy balance/conservation between the dust emission and attenuated stellar emission. We model this independently for each SFH time step by requiring the total integrated IR luminosity ($L_{\rm TIR}$) from dust emission to be equal to the total integrated absorbed stellar luminosity ($L_{\star}^{\rm abs}$). Assuming azimuthal symmetry, this is given by
\begin{equation} \label{eq:GeneralLTIR}
   L_{\rm TIR} = L_{\star}^{\rm abs} = 2 \pi d_L\!^2 \int^{\pi}_0 \int^{\infty}_0 \Big(F_{\nu}\!^{\rm unatt}-F_{\nu}\!^{\rm att} \Big) \sin\theta \, d\nu \, d\theta,
\end{equation}
where $F_{\nu}\!^{\rm unatt}$ and $F_{\nu}\!^{\rm att}$ are, respectively, the unattenuated and attenuated fluxes from the stellar emission. For an inclination-independent attenuation curve, this simplifies to the energy balance assumption:
\begin{equation}
    \label{eq:IncIndepLTIR}
    L_{\rm TIR} =  L_{\star}^{\rm abs} = L_{\rm bol}^{\rm unatt} - L_{\rm bol}^{\rm att},
\end{equation}
where $L_{\rm bol}^{\rm unatt}$ is the bolometric luminosity of the stellar population without attenuation being applied, and $L_{\rm bol}^{\rm att}$ is the bolometric luminosity after attenuation is applied assuming the line-of-sight emission is isotropic.

However, when using our inclination-dependent attenuation curves that assume anisotropic emission, Equation~\ref{eq:GeneralLTIR} does not simplify as easily, since $F_{\nu}\!^{\rm att}$ is a function of inclination (or $\theta$). To compute $L_{\star}^{\rm abs}$, the polar angle $\theta$ in Equation~\ref{eq:GeneralLTIR} can be replaced with inclination  and simplified to
\begin{equation}
    \label{eq:IncDepLTIR}
   L_{\rm TIR} =  L_{\star}^{\rm abs} = L_{\rm bol}^{\rm unatt} - \int^{\pi/2}_{0} \tilde{L}_{\rm bol}^{\rm att} (i) \sin i \, di,
\end{equation}
where $\tilde{L}_{\rm bol}^{\rm att}(i) \equiv 4 \pi d_L\!^2 \int^{\infty}_0 F_{\nu}\!^{\rm att}(i) \, d\nu$ and
\begin{equation}
    \label{eq:LbolEsc}
   L_{\rm bol}^{\rm att} = \int^{\pi/2}_{0} \tilde{L}_{\rm bol}^{\rm att}(i) \sin i \, di.
\end{equation}
Therefore, Equation~\ref{eq:LbolEsc} must be integrated over inclination to generate $L_{\rm bol}^{\rm att}$ so that the $L_{\star}^{\rm abs}$ can be calculated for the inclination-dependent model.

To calculate $L_{\star}^{\rm abs}$, we numerically integrated Equation~\ref{eq:LbolEsc} using the trapezoidal method for a grid of inclination angles spanning 0 to $\pi/2$. Due to $\tilde{L}_{\rm bol}^{\rm esc}$ being determined from the inclination-dependent attenuation curve, the attenuation had to be computed for this grid of inclination angles along with the input inclination. Rather than computing this integral and attenuation multiple times for each galaxy in our sample while fitting an SED, we precomputed an array of $L_{\star}^{\rm abs}$ for each SFH time step once from Equation~\ref{eq:IncDepLTIR} using a fine grid of the inclination-dependent attenuation parameters in Equation \ref{eq:TuffTotAtt} (i.e., $i$, $\tau_B^f$, $F$, $r^{0,\rm disk}$, and $r^{0,\rm bulge}$). This fine grid consisted of 51 equally spaced grid points for each attenuation parameter, except inclination. We used 70 inclination angles to ensure an accurate calculation of the integral. We also added 10 additional finely spaced grid points to $\tau_B^f$ between 0 and 0.1 (i.e., 0.01--0.1 in steps of 0.01) to ensure the accuracy of the $L_{\star}^{\rm abs}$ array, due to these values not being in the original \citet{2004A&A...419..821T} tabulations. The $L_{\star}^{\rm abs}$ of the last two SFH time steps had to be computed for a grid of redshifts, since the age range of the step varied with the redshift, as described in Section~\ref{sec:SEDFitPro}. The redshift grid was computed in steps of 0.01, since this was the accuracy used for our spectroscopic redshifts. We then linearly interpolated between the fine attenuation parameter grid points to determine $L_{\star}^{\rm abs}$ for any possible combination of attenuation parameters at a given redshift. Comparing the interpolated $L_{\star}^{\rm abs}$ values from the precomputed arrays to $L_{\star}^{\rm abs}$ values computed from the exact attenuation parameters and 70 inclination grid points using Equation~\ref{eq:IncDepLTIR} showed that the interpolated values were always within 0.5\% of the exact calculations of $L_{\star}^{\rm abs}$.

We recommend that if a precomputed array of $L_{\star}^{\rm abs}$ is not used, a grid of inclinations should be used that minimizes the computational time and maximizes the accuracy of the integral. We have allowed for this possibility in \texttt{Lightning} and provided the optimal grid, if one is not supplied. To determine the optimal grid, we computed the integral for grids of 3 to 70 equally spaced inclination angles for various combinations of attenuation curve input parameters. We found that using $\gtrsim13$ grid points for the integral resulted in $\lesssim0.5\%$ difference in $L_{\star}^{\rm abs}$ compared to the grid with 70 points. Using more points minimally changed this difference, and fewer points rapidly increased the difference. Therefore, when computing the integral in Equation~\ref{eq:LbolEsc} without a specified grid of inclinations, we required 13 equally spaced inclinations besides the input inclination. We recommend using a precomputed array of model $L_{\star}^{\rm abs}$ rather than calculating it with the optimal grid for each new combination of attenuation parameters. Excluding the time required to make the precomputed array, using it is approximately 10 times faster computationally per calculation of $L_{\star}^{\rm abs}$ than using the optimal grid.

\section{SED Fitting Results} \label{sec:Results}

\subsection{Inclination-independent Comparison Fits} \label{sec:IncIndFits}

To test the efficacy of the inclination-dependent attenuation prescription, we derived SFHs using the inclination-independent \citet{2000ApJ...533..682C} attenuation curve in its original form for comparison. We used this attenuation curve within our adaptive MCMC procedure along with energy balance and our \citet{2007ApJ...657..810D} dust model. The \citet{2000ApJ...533..682C}  attenuation curve was chosen due to its widespread use in SED fitting of deep-field galaxies \citep[e.g.,][]{2005ApJ...626..680D,2010ApJ...709..644I,2014ApJS..214...24S,2015ApJ...808..101M}.

In order to reduce potential degeneracies in the dust model, we set the parameters $U_{\rm max}=3\times10^5$ and $\alpha=2$ as discussed in Section \ref{sec:Dust}. We also limit the dust models to be of MW composition with uniform priors spanning $0.4\% \le q_{\rm PAH} \le 4.6\%$ and $0.7 \le U_{\rm min} \le 25$. This range and set of fixed parameters is the ``restricted'' dust model recommended by \citet{2007ApJ...663..866D} when submillimeter data are unavailable. The range of $q_{\rm PAH}$ spans the full range of values for the MW composition; however, the lower limit of $U_{\rm min}$ has been chosen to be 0.7 instead of 0.1. This is because small values of $U_{\rm min}$ correspond to cold dust temperatures, which require rest-frame submillimeter data ($\lambda_{\rm rest} > 500$~$\mu$m) to be properly constrained.

Besides the degeneracies in the dust model, the other main degeneracy in our fits is the well-established age-reddening-metallicity degeneracy. To help minimize this, we fixed the metallicity to the solar value for all of our age bins. We note that this ignores the underlying metallicity evolution and could cause systematic variation in our SFHs and stellar mass estimates. As metallicity decreases, the intrinsic UV-optical emission for our models increases for a fixed SFR. This can lead to slightly decreased SFRs for the younger populations of the SFH, assuming fixed attenuation, due to the younger populations dominating the UV-optical emission. However, the stellar mass estimates would be relatively unaffected due to the older populations, which mainly emit at wavelengths in the NIR and minimally contribute to the UV-optical emission, most strongly affecting the mass estimates. Further, fixing the metallicity still leaves some age-reddening degeneracy, but this is reduced by our energy balance assumption (see Section \ref{sec:EnCon}). Therefore, we do not expect any material impact on our results by ignoring metallicity evolution.

\begin{figure*}[t!]
\centerline{
\includegraphics[width=18cm]{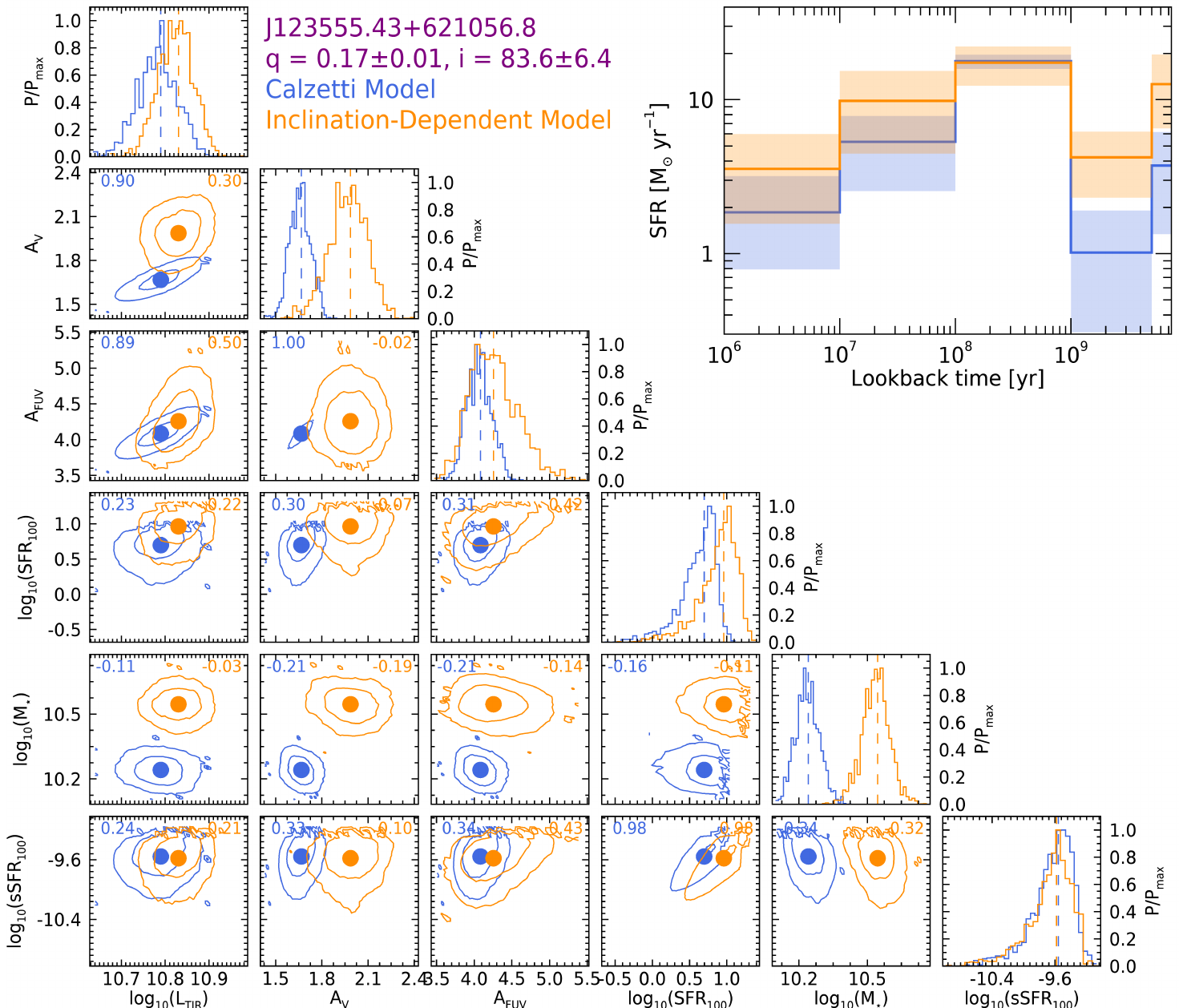}
}
\caption{
(\textit{Lower left triangle plot}): Probability distribution functions in terms of $P/P_{\rm max}$ (diagonal elements) and the 68\% and 95\% confidence contours for $L_{\rm{TIR}} \ [L_\odot]$ , $A_V \ [{\rm mag}]$, $A_{\rm FUV} \ [{\rm mag}]$, $\rm{SFR}_{100} \ [M_\odot \ yr^{-1}]$, $M_{\star} \ [M_\odot]$, and $\rm{sSFR}_{100} \ [yr^{-1}]$ parameter pairs (off-diagonal elements) for our most inclined example galaxy, J123555.43+621056.8. This galaxy is the galaxy with a purple outline in Figure~\ref{fig:postagestamp}. The vertical dashed lines in the histograms and solid colored circles in the contour plots indicate the median values of each parameter. The Pearson correlation coefficients for each set of parameters are shown in the upper corners of each contour plot. Blue represents the results from the \citet{2000ApJ...533..682C} fits, and orange represents the results from the inclination-dependent fits with the image-based inclination prior. (\textit{Upper right corner}): The resulting median SFH and 16\%--84\% uncertainty range for J123555.43+621056.8 with the same color scheme. The youngest age bin lower bound has been truncated to $10^6$ yr for plotting purposes, but truly extends to 0 yr. 
}
\label{fig:tridistplot}
\end{figure*}

With our adopted priors on the dust model, we ran the adaptive MCMC algorithm for $10^5$ iterations for an initial fit on each galaxy's SED with arbitrarily chosen starting values. To test for convergence to a single best solution, we ran 10 parallel chains at random starting values between 0 and 10 $M_{\odot}$~yr$^{-1}$ for the five SFH bins and random starting values within the attenuation and dust parameter ranges. We chose the starting range for the SFH bins based off of the initial fits' SFH distributions, of which 75\% had values less than 10 $M_{\odot}$~yr$^{-1}$. A larger starting range could result in a drastically increased burn-in phase if a starting value was much larger than the solution. To confirm the convergence of the parallel chains, we performed the Gelman-Rubin test \citep{Gelman1992,Brooks1998} on the last 5000 iterations of the chains. This test indicated that all chains for each galaxy converged to the same solution by the final 5000 iterations (i.e., $\sqrt{\hat{R}} \approx 1$). Therefore, we used the last 5000 iterations of the parallel chain that produced the minimum median $\chi^2$ for our parameter distributions and subsequent analysis. To test the quality of fits to the SEDs, we performed a $\chi^2$ goodness of fit test using the minimum $\chi^2$ of each galaxy's chain. The resulting distribution of $P_{\rm null}$ from this test showed a relatively flat distribution (i.e., expected distribution of $\chi^2$). Therefore, we conclude that the \cite{2000ApJ...533..682C} model can  acceptably model the SEDs.

\begin{figure*}[t!]
\centerline{
\includegraphics[width=18cm]{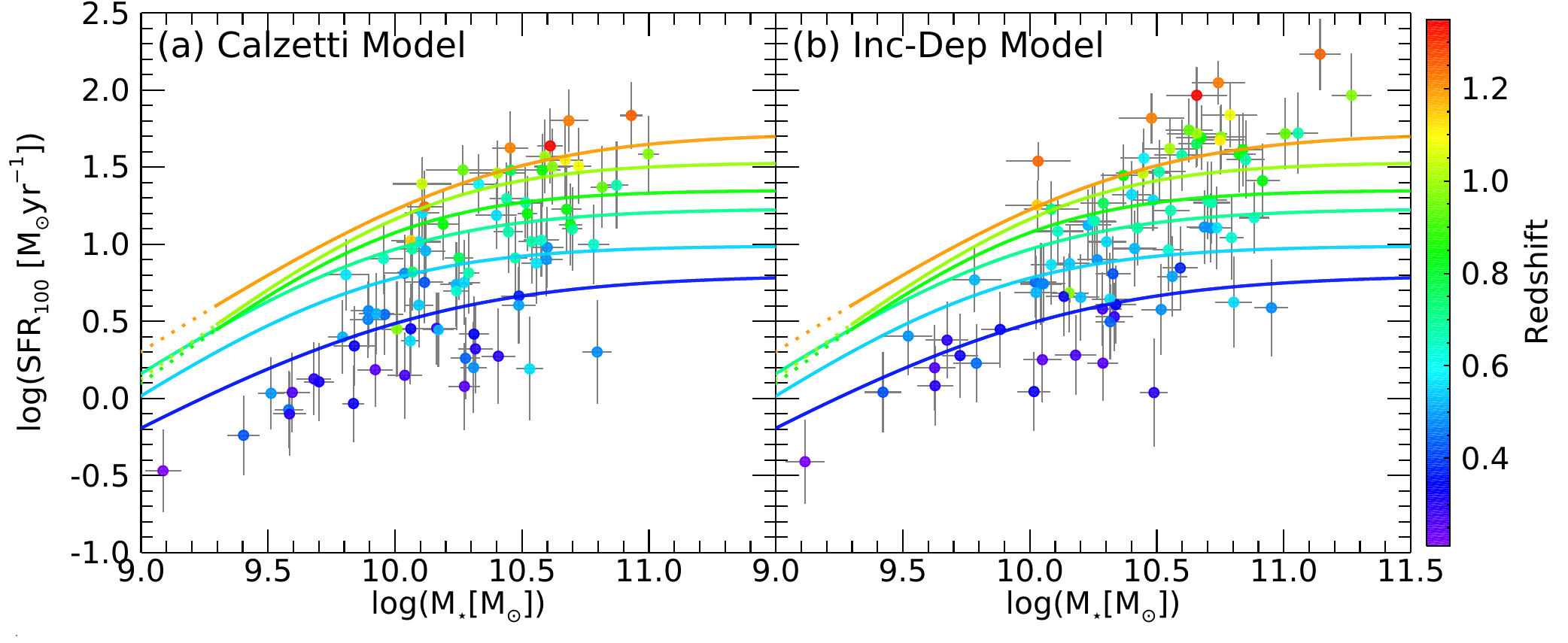}
}
\caption{
$\rm{SFR}_{100}$ vs. $M_\star$ for the 82 galaxies in our final sample colored by redshift. The colored curves show the location of the star-forming galaxy MS as derived by \citet{2015ApJ...801...80L} at their median redshifts given by the corresponding color in the color bar. The dotted sections of the higher redshift curves show the extrapolated region of the curves beyond the lower stellar mass limits. Panel (\textit{a}) shows the $\rm{SFR}_{100}$ and stellar masses derived from the \citet{2000ApJ...533..682C} model, and panel (\textit{b}) shows the values derived from the inclination-dependent model with the image-based inclination prior. Both panels show that most galaxies in the final sample tend to follow the star-forming MS at low redshifts ($z<0.7$). As for higher redshift galaxies ($z>0.7$), the inclination-dependent model shows that galaxies tend to be above the MS, while the \citet{2000ApJ...533..682C} model shows that they tend to follow the MS.
}
\label{fig:sfms}
\end{figure*}

An example of the distributions for the parameters of interest, which are $L_{\rm TIR}$, $V$-band attenuation ($A_V$), FUV-band attenuation ($A_{\rm FUV}$), recent average SFR of the last 100 Myr ($\rm{SFR}_{100}$), total stellar mass ($M_{\star}$), and specific SFR of the last 100 Myr ($\rm{sSFR}_{100}$), are shown in Figure \ref{fig:tridistplot} as the blue lines for our most inclined example galaxy, J123555.43+621056.8. The resulting median SFH and its 16\%--84\% uncertainty range is also shown in the upper right corner. In Figure~\ref{fig:sfms}(a), we show how the derived $M_{\star}$ and $\rm{SFR}_{100}$ from these fits compare to the star forming galaxy main sequence (MS) from \citet{2015ApJ...801...80L}. The results from these fits tend to follow the MS at their respective redshift. Additional diagnostic plots showing the free parameter distributions and the global trends for all galaxies in the sample can be found in Appendix~\ref{sec:DiagFigs}.

\subsection{Inclination-dependent Fits} \label{sec:IncDepFits}

For our inclination-dependent fits, we used our adaptive MCMC procedure with energy conservation, the ``restricted'' \citet{2007ApJ...657..810D} dust model, and the inclination-dependent attenuation curves. For the inclination-dependent attenuation curves, we fix $r^{0,\rm old}=0$ for the first three age bins of our SFHs and $r^{0,\rm old}=1$ for the older two age bins, as to define them as the young and old populations, respectively, as discussed in Section~\ref{sec:AttCurve}. The third age bin (0.1--1~Gyr) is considered a ``young'' age bin due to the requirement that all age bins that contain ages $<$ 500~Myr must be considered part of the young population as to have their nonnegligible UV emission attenuated.\footnote{We tested how the choice of this third age bin upper limit affects our results and found that changing the upper limit to 500~Myr or 1.5~Gyr had no statistical impact on the results (see Section~\ref{sec:Discussion} and Figure~\ref{fig:incvproperty}).} Further, as stated in Section~\ref{sec:SampleSelect}, we only analyzed SEDs of disk-dominated galaxies, rather than disk galaxies in general. Since we selected disk-dominated galaxies with approximately no bulge, we set $B/D=0$ to reduce the number of free parameters and potential degeneracies. As stated by \citet{2004A&A...419..821T} and noted in Section~\ref{sec:AttCurve}, increasing $B/D$ with $\tau_B^f$ constant can have the same effect on the attenuation curve as increasing $\tau_B^f$ for a ``pure'' disk (i.e., $B/D=0$). We therefore remove this degeneracy by selecting our sample to be disk-dominated, or as close to being a ``pure'' disk as possible. We note, however, that the presence of a small bulge has the effect of systematically increasing the derived values of $\tau_B^f$. In addition to this model degeneracy, there is another possible degeneracy between inclination and $\tau_B^f$. As discussed in Section~\ref{sec:AttCurve}, increasing the inclination or $\tau_B^f$ has the effect of steepening the attenuation curve. We discuss how this degeneracy affects the derived inclinations in Section~\ref{sec:SEDIncEst}.

\begin{figure*}[t!]
\centerline{
\includegraphics[width=18cm]{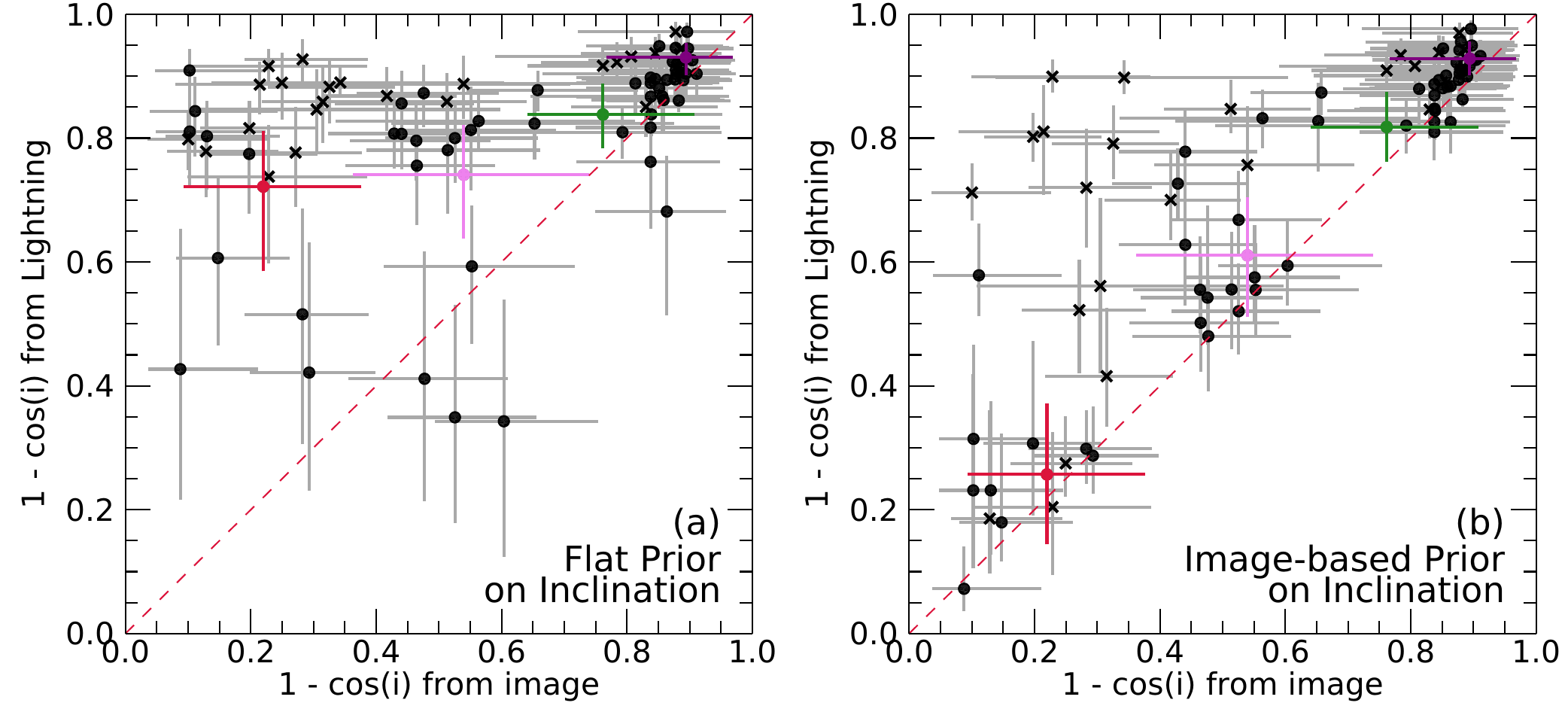}
}
\caption{
(\textit{a}) Inclination ($1-\cos i$) derived from \texttt{Lightning} with flat inclination prior. (\textit{b}) Inclination derived from \texttt{Lightning} with image-based inclination prior. Both are vs. the inclination derived from the image-based Monte Carlo simulation. The error bars span the 16th and 84th percentiles rather than 1$\sigma$ due to the asymmetry of the image-based inclination distributions. The colored points are the inclination estimates of the four example galaxies shown in Figure~\ref{fig:postagestamp} using the same color as the outline of the corresponding postage stamp. Red, pink, green, and purple represent J123626.62+621252.1, J123654.99+621658.5, J033231.18-274017.5, J123555.43+621056.8, respectively. Using the image-based priors aligns more galaxies to the unity line. Galaxies that still remain off of the line tend to have higher $L_{\rm TIR}$ and $z \geq 0.7$. In both panels, galaxies with $z < 0.7$ are indicated by the circles, and those with $z\geq0.7$ are indicated by the X marks.
}
\label{fig:incvinc}
\end{figure*}

Beyond these degeneracies, we note that certain parameters could theoretically be linked together to make an even more physically motivated model. For example, the attenuation from the clumpy birth cloud component, $F$, could be linked to the fraction of the total dust luminosity that is radiated by dust grains in regions where $U >10^2$, or $f_{\rm PDR}(U_{\rm min}, U_{\rm max}, \gamma_{\rm dust})$ (given by Equation 29 in \citealt{2007ApJ...657..810D}), which is typically associated with photodissociation regions (PDRs) near newly born luminous stars \citep{2007ApJ...657..810D}. Not considering this linkage could result in nonphysical results where $F$ is high and $f_{\rm PDR}$ is low. However, implementing potential linkages between parameters like this is beyond the scope of this paper, but is something that could be explored in future work.

For these fits, we ran the adaptive MCMC algorithm for $2\times10^5$ iterations. A larger number of iterations here compared with the \citet{2000ApJ...533..682C} fits in Section~\ref{sec:IncIndFits} was required due to the larger parameter space so that the best solution could be reached. We again tested for convergence of the chains to a single best solution by running 10 parallel chains at random starting values between 0 and 10 $M_{\odot}$~yr$^{-1}$ for the five SFH bins and random starting values within the attenuation and dust parameter ranges. The Gelman-Rubin test was then performed on the last 5000 iterations of the parallel chains, which indicated that convergence to the same solution had been achieved by the final 5000 iterations. Therefore, like the \citet{2000ApJ...533..682C} fits, we used the last 5000 iterations of the parallel chain that had the minimum median $\chi^2$ for our parameter distributions.

We tested the quality of these fits by performing a $\chi^2$ goodness of fit test using the minimum $\chi^2$ of each galaxy's chain. This test showed that the resulting distribution of $P_{\rm null}$ had a relatively flat distribution (i.e., expected distribution of $\chi^2$). Therefore, we concluded that our inclination-dependent model can also acceptably model these SEDs.

\subsection{SED Inclination Estimates} \label{sec:SEDIncEst}

After fitting the SEDs with the inclination-dependent model, we compared the derived inclination PDFs from the fits to the inclination PDFs from the image-based Monte Carlo simulation described in Section~\ref{sec:GalInc}. This was done to determine the predictive power of the inclination-dependent model for inclination with the presence of the inclination-$\tau_B^f$ degeneracy. Figure~\ref{fig:incvinc}(a) shows this comparison as the median values from each distribution and the 16th and 84th percentile error ranges. This shows that \texttt{Lightning} tends to favor solutions at high inclinations, with a median value never falling below $1-\cos i \approx0.3$, while the image-based method has inclinations down to $1-\cos i \approx0.1$. To test the consistency of the fits' inclination PDFs with the image-based inclination PDFs, we computed $R$, which we define as the ratio of the intersection area to the union area of the two distributions, for each galaxy. This method would result in $R=1$ if the two distributions were identical and $R=0$ if they had no overlap. Using these ratios, we chose to set a value of $R_{\rm cutoff}=0.05$ as the cutoff at which we define $R$ values lower than this cutoff to have inclinations that are in disagreement between methods. For these fits, 60 out of the 82 PDFs ($\approx$73\%) had $R>R_{\rm cutoff}$ with a median of $R=0.29$.

Due to this relatively large disagreement ($\approx$27\%) in inclination estimates and the apparent bias of the fit inclinations to higher values, we decided to refit the SEDs using the image-based PDFs of inclination as priors to minimize the inclination-$\tau_B^f$ degeneracy and to force the predicted inclinations to be more consistent with the image-based estimates. The method for refitting these SEDs and testing for convergence of the Markov chains was exactly the same as in Section~\ref{sec:IncDepFits}, except for the introduction of the new prior on inclination. All other parameters were still fit using flat priors. Convergence of these chains to a single solution was achieved by the final 5000 iterations. We then used the last 5000 iterations selected using the same method described above to make our final parameter distributions. Testing the quality of these fits with a $\chi^2$ goodness of fit test showed again that the resulting distribution of $P_{\rm null}$ had a relatively flat distribution (i.e., expected distribution of $\chi^2$). Therefore, we concluded that adding the image-based inclination priors had no effect on the acceptability of the model, and we adopted these fits as our inclination-dependent fits for all further analyses.

Example distributions for the parameters of interest for our example galaxy, J123555.43+621056.8, from the inclination-dependent fits using the image-based prior are shown in Figure \ref{fig:tridistplot} as the orange lines. Comparing these distributions to the distributions from the \citet{2000ApJ...533..682C} fits shows that most parameters are highly consistent between models with the exception of $A_V$ and $M_{\star}$. These inconsistencies and how they vary with inclination will be discussed in Section \ref{sec:Discussion}. As for the SFH in the upper right corner, the inclination-dependent model predicts higher median SFR at all but the third age bin. However, these values are consistent between models when considering the uncertainty. In Figure~\ref{fig:sfms}(b), we show how the derived $M_{\star}$ and $\rm{SFR}_{100}$ from these inclination-dependent fits compare to the star forming galaxy MS from \citet{2015ApJ...801...80L}. The results from these fits tend to follow the MS for galaxies with $z\lesssim0.8$. However, galaxies with $z\gtrsim0.8$ tend to fall above the MS, and we discuss the potential causes for this below.

We then compared our inclinations from the updated fits with inclination priors to the image-based inclinations to determine the statistical impact of the prior. Figure~\ref{fig:incvinc}(b) shows that indeed the inclinations for many of the galaxies were influenced by the use of the prior. To quantitatively test this impact, we again computed $R$ for each galaxy for the updated fits and image-based PDFs. For these fits, 72 out of the 82 PDFs ($\approx$88\%) had $R>R_{\rm cutoff}$ with a median of $R=0.39$, which is an increase in the number of galaxies by 15\% and median $R$ by 0.10. This increase in agreement and median $R$ showed that the inclination priors were informative for several galaxies and that adding the image-based priors allowed for more consistent inclination distributions between methods.

\begin{figure*}[t!]
\centerline{
\includegraphics[width=18cm]{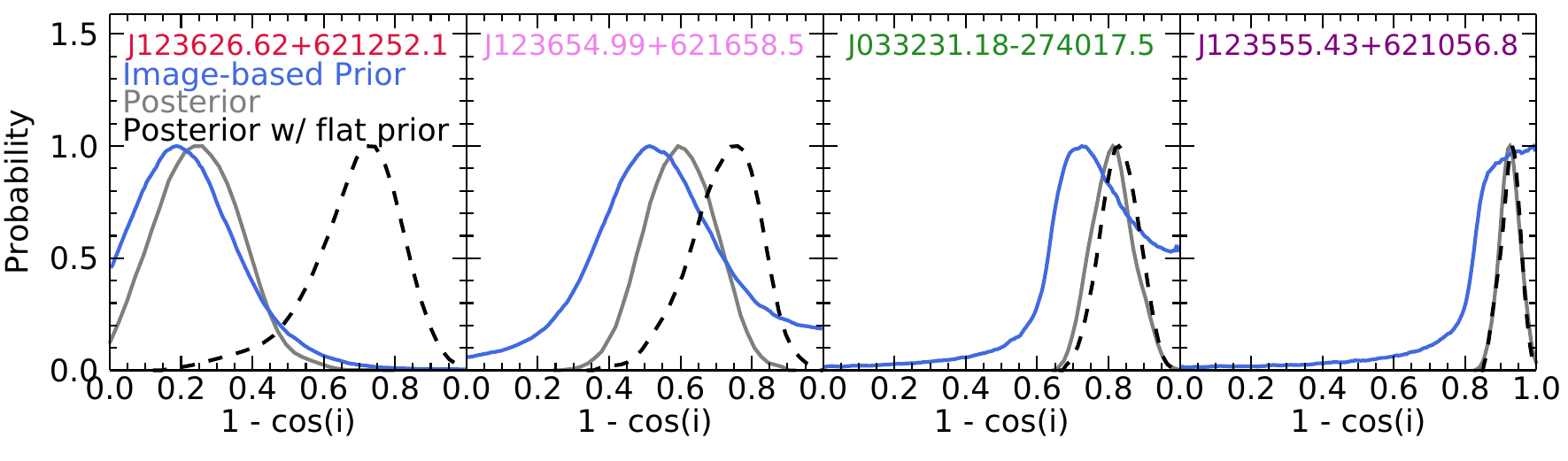}
}
\caption{
Image-based prior and resulting posterior probability distributions of the inclination ($1-\cos i$) as the blue and gray lines, respectively, as well as the resulting posterior assuming a flat prior as the black dashed line for the four example galaxies. Each distribution is normalized to 1 for comparison purposes. The names of the galaxies are colored using the same color as the outline of the corresponding postage stamp in Figure~\ref{fig:postagestamp}. For low-inclination galaxies like J123626.62+621252.1 and J123654.99+621658.5, the image-based priors are often informative, while for high-inclination galaxies like J033231.18-274017.5 and J123555.43+621056.8, inclination is primarily constrained by the likelihood.
}
\label{fig:galincpdf}
\end{figure*}

Examples of the prior and resulting posterior probability distributions from these updated fits can be seen in Figure~\ref{fig:galincpdf} for the four example galaxies as the blue and gray lines, respectively. The black dashed lines show the posteriors from the fits with the flat inclination prior. In some cases, the image-based priors are informative (e.g., J123626.62+621252.1 and J123654.99+621658.5), while in other cases they are not (e.g., J033231.18-274017.5 and J123555.43+621056.8).

As for the galaxies still with $R<R_{\rm cutoff}$, adding the image-based priors only had a slight effect, with the median $R$ increasing from $R=0.01$ to $R=0.02$. Due to this inconsistency, even after adding the image-based priors, we further inspected these galaxies to determine the potential source of this inconsistency. We initially checked for visual morphological differences in the sample, and the galaxies that had $R<R_{\rm cutoff}$ tended to have bright, blue, off-center star forming clumps. To quantify this observed difference for each galaxy, we measured the concentration ($C$), asymmetry ($A$), and clumpiness ($S$) morphology parameters following the methods of \citet{2004AJ....128..163L} for the HST/ACS F435W postage stamp images. However, $S$ was deemed to be an unreliable metric, due to the large range in redshift of our sample, which causes a large range in the physical resolution of each galaxy's postage stamp as well as decreasing signal-to-noise ratio. Therefore, we measured the second-order moment of the brightest 20\% of the galaxy's flux ($M_{20}$) as defined in \citet{2004AJ....128..163L}, which also measures the clumpiness of a galaxy. This metric is influenced less by the variation in the signal-to-noise ratio compared to $S$ (see Figure 5 in \citealt{2004AJ....128..163L}), and would therefore be a more reliable metric with this variation in redshift. Comparing these parameters for the galaxies with $R>R_{\rm cutoff}$ to those with $R<R_{\rm cutoff}$, we found slightly lower values of $C$ and higher values of $A$ and $M_{20}$ for the galaxies with $R<R_{\rm cutoff}$, which implies off-center clumps could be present more often in these objects. However, a two-sided KS test showed that these differences are not statistically significant ($p\textrm{-value}>0.5$), and therefore, we could not confidently conclude that morphological differences are the driving factor for this disagreement in inclination.

Another possibility, in addition to morphology, that could be responsible for the disagreement in inclinations is limitations in the SED fitting techniques. As seen in Figure~\ref{fig:incvinc}(b), if \texttt{Lightning} does not predict an inclination in agreement with the prior image-based inclination, it typically predicts an inclination value higher than the peak of this prior distribution. This is driven by the data requiring a relatively high attenuation made available by high inclinations models (and higher $\tau_B^f$). This high attenuation requirement comes from an elevated $L_{\rm TIR}$ and the energy conservation requirement. Comparing $L_{\rm TIR}$ of the galaxies with $R>R_{\rm cutoff}$ and $R<R_{\rm cutoff}$, the galaxies with $R<R_{\rm cutoff}$ had a larger median $L_{\rm TIR}$ by a factor of $\approx$6 over the galaxies with $R>R_{\rm cutoff}$. A 2D KS test showed that this difference was highly significant ($p\textrm{-value}<10^{-5}$), and likely a driving factor for this disagreement in inclination. This disparity in $L_{\rm TIR}$, which is also present in the \citet{2000ApJ...533..682C} fits, could arise because either the dust emission is actually elevated, or the dust emission is poorly constrained for these galaxies. If the dust emission is truly elevated, it could be that these galaxies are low-luminosity or obscured AGNs that made it though our removal of AGNs in the sample selection process, since AGNs are known to have increased $L_{\rm TIR}$ compared to star forming galaxies \citep{2012ApJ...759..139K}. However, testing to see if the dust emission is truly elevated would require additional IR data to fully constrain the dust emission of these galaxies.

\begin{figure}[t!]
\centerline{
\includegraphics[width=8.5cm]{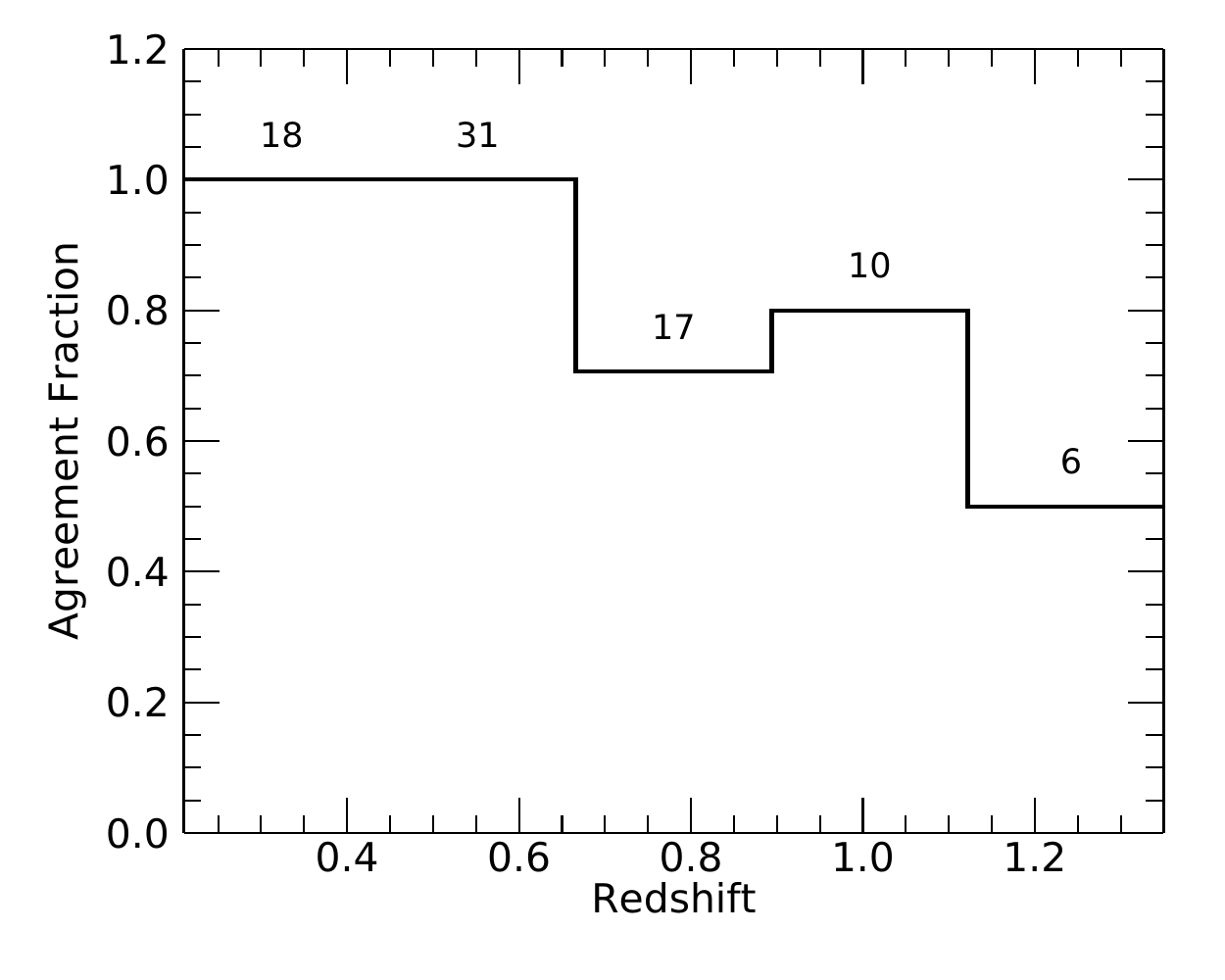}
}
\caption{
The fraction of galaxies with inclinations agreeing between the image-based and SED-based (with the image-based inclination as a prior) inclinations vs. redshift. Agreement was defined as $R>R_{\rm cutoff}$ where $R_{\rm cutoff}=0.05$. The numbers of galaxies that are contained within each redshift bin are shown above that respective bin. 
}
\label{fig:agreefrac}
\end{figure}

To further check for limitations of the SED fitting techniques, we compared the redshifts of the galaxies with $R>R_{\rm cutoff}$ to those with $R<R_{\rm cutoff}$. The agreement fraction versus the redshift is displayed in Figure~\ref{fig:agreefrac}, with the total number of galaxies within each redshift bin labeled above the respective bin. The agreement fraction is defined as the number of galaxies with $R>R_{\rm cutoff}$ divided by the total number of galaxies within the respective redshift bin. From this, it can be seen that as the redshift increases the agreement fraction decreases, with a drop-off in the level of agreement above $z\approx0.7$. A 2D KS test showed that this redshift variation was highly significant with $p\textrm{-value}<10^{-3}$. It is possible that this variation and drop-off at $z\approx0.7$ is due to the \citet{2004A&A...419..821T} attenuation curves potentially not being physically appropriate to model these galaxies. The curves were made from the known thin and thick disk structure of local galaxies. However, it has been shown that galaxies with $z\gtrsim1$ tend to be significantly thicker and dynamically hotter than galaxies in the local universe \citep[e.g.,][]{2013ApJ...773...43B,2014ApJ...792L...6V,2017ApJ...847...14E,2019MNRAS.490.3196P,2019MNRAS.484.5170Z}. This would explain the drop-off in agreement at $z\sim1$, the elevated $L_{\rm TIR}$ for the galaxies with $R<R_{\rm cutoff}$ due to the dynamically hotter environment, and would tie into potential morphological differences. However, to confirm this, we would need more sources at these higher redshifts in order to have better statistics.

\section{Discussion} \label{sec:Discussion}

In Section~\ref{sec:Results}, we were able to acceptably fit the SEDs of our sample galaxies with \texttt{Lightning} using both the \citet{2000ApJ...533..682C} and inclination-dependent attenuation models. Since the fits are independent of each other and the only difference in the models is the attenuation curves, we were able to directly compare these fits to determine the effect of incorporating inclination on their derived SFHs. 
However, as discussed in Section~\ref{sec:SEDIncEst}, the inclination-dependent model has a decreasing agreement between the image-based and SED-based inclination estimates with the redshift, suggesting some redshift evolution effects influence these higher redshift fits. Therefore, we chose to limit our comparisons in this section to the 58 galaxies that have $z<0.7$ to mitigate any redshift evolution effects present in the inclination-dependent fits.

We first compared the values of $L_{\rm TIR}$ between fits to ensure the estimated global dust-absorbed radiative power was comparable between models. Due to the dust model being of the same form \citep[i.e.,][]{2007ApJ...657..810D} for both fits, $L_{\rm TIR}$ should, in principle, be similar between the \citet{2000ApJ...533..682C} and inclination-dependent fits. If significant differences in $L_{\rm TIR}$ occurred, then reliable comparisons between stellar properties would not be meaningful. This is due to the energy balance/conservation requirement, which tied the total amount of attenuation to $L_{\rm TIR}$ as discussed in Section~\ref{sec:EnCon}. Therefore, differences in $L_{\rm TIR}$ between fits would result in differences in the total attenuation between fits. These differences would boost the values of the intrinsic stellar properties for the fit with an elevated total attenuation, and potentially obscure any differences in stellar properties between fits that reveal trends with the inclination.

\begin{figure}[t!]
\centerline{
\includegraphics[width=8.5cm]{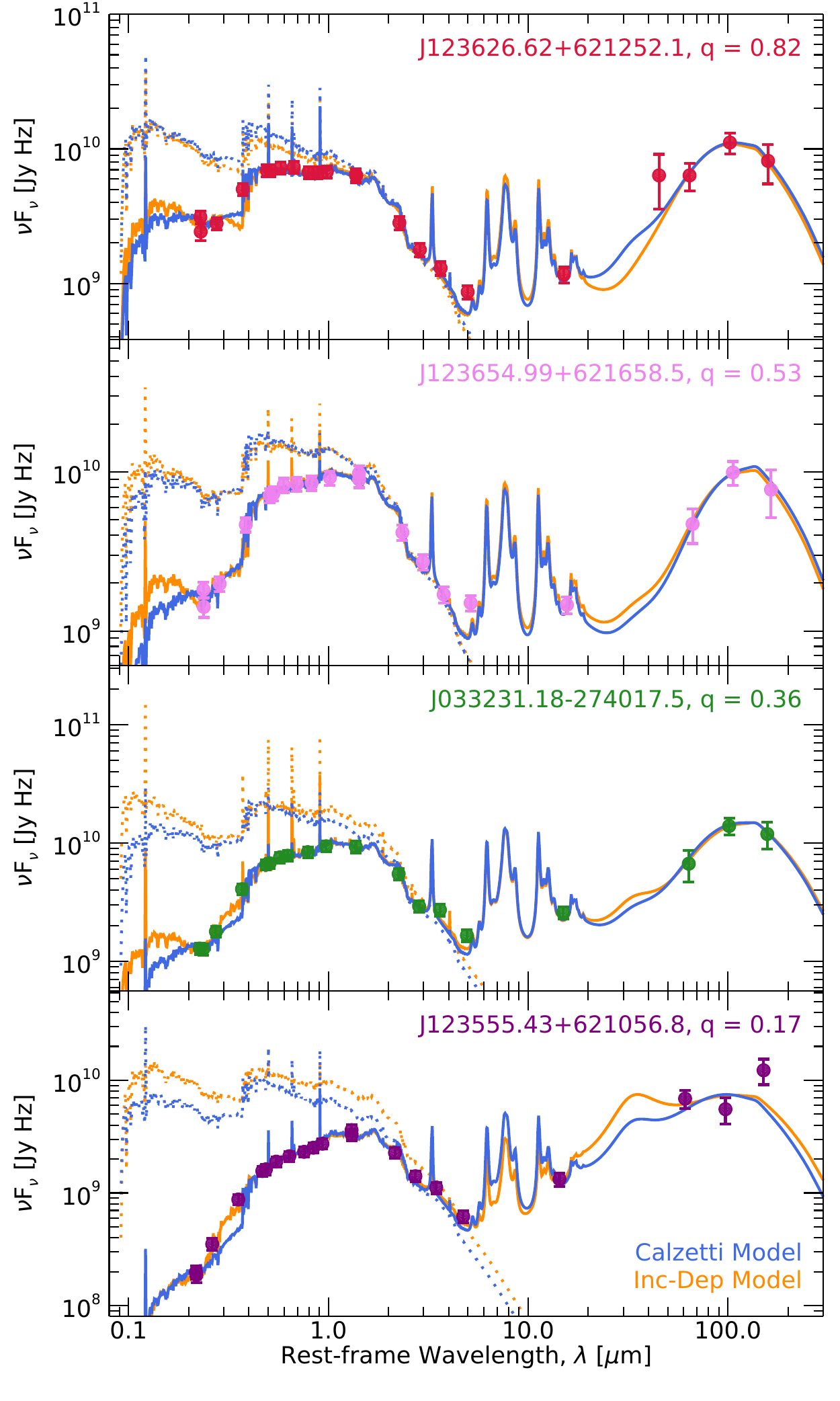}
}
\caption{
The best-fit models to the broadband SEDs from the \citet{2000ApJ...533..682C} fits and inclination-dependent fits with the image-based inclination prior for the four example galaxies shown in Figure~\ref{fig:postagestamp} as the blue and orange lines, respectively. The solid lines are the combined dust and attenuated stellar models, and the dotted lines are the unattenuated stellar models. The broadband SEDs and the names of the galaxies are colored using the same color as the outline of the corresponding postage stamp in Figure~\ref{fig:postagestamp}. The galaxies are arranged from top to bottom from the least inclined to the most inclined. 
}
\label{fig:obsmodsed}
\end{figure}
\begin{figure*}[t!]
\centerline{
\includegraphics[width=18cm]{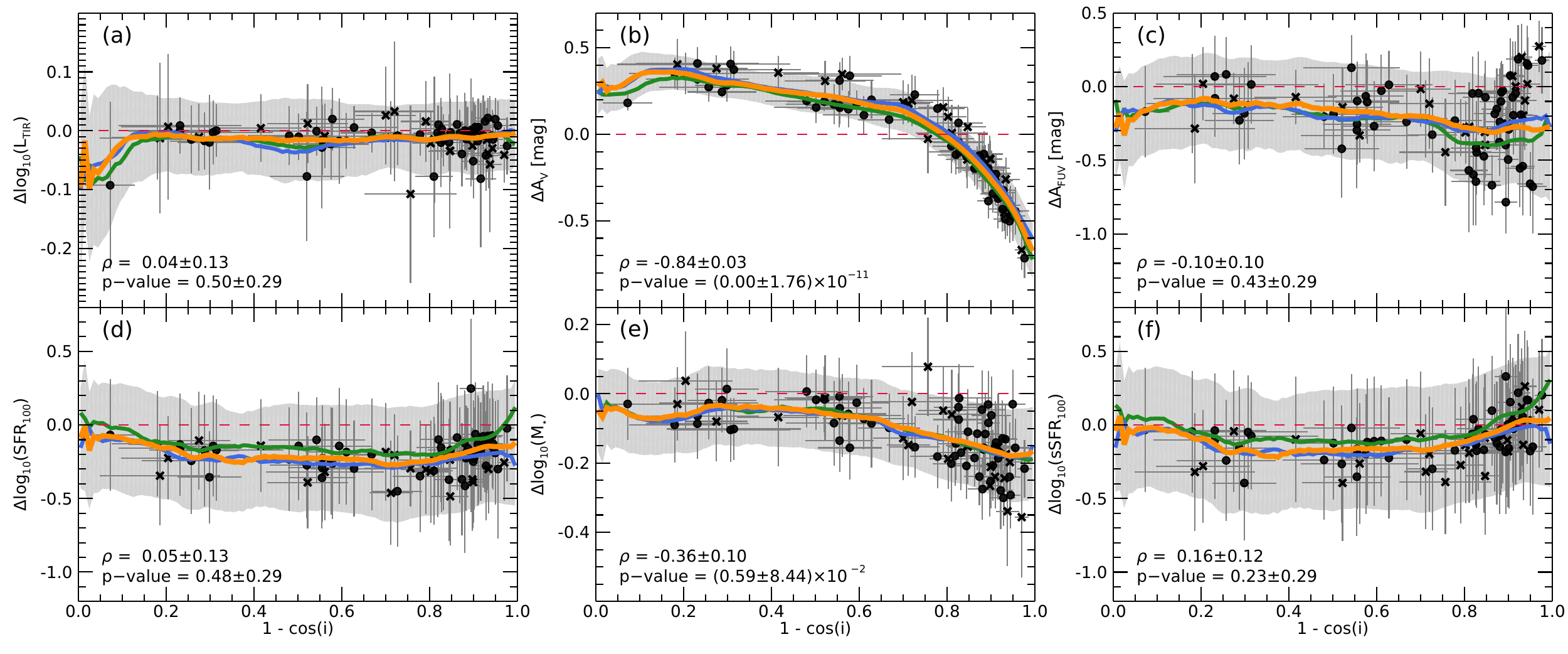}
}
\caption{
The panels show the median and 1$\sigma$ dispersion of the logarithmic difference ($\log_{10}(\rm{property_{Calz}})-\log_{10}(\rm{property_{Dep}})$; panels (\textit{a}), (\textit{d}), (\textit{e}), and (\textit{f})) and difference ($\rm{property_{Calz}}-\rm{property_{Dep}}$; panels (\textit{b}) and (\textit{c})) between the \citet{2000ApJ...533..682C} and inclination-dependent fits for the parameters of interest vs. the inclination ($1-\cos i$) derived from the SED fittings. The solid circles represent the galaxies with $z<0.7$, and the X's represent the galaxies with $z\geq0.7$, which are not used in deriving the trend lines or correlations. The orange line and light gray 1$\sigma$ dispersion range are the mean and standard deviation of $\Delta \cos i = 0.01$ bins calculated by using all 5000 elements within the respective distributions of each property as data points. The green and blue lines are the mean of $\Delta \cos i = 0.01$ bins for fits where the third age bin upper bound is adjusted to 500~Myr and 1.5~Gyr, respectively. These fits show no significant differences from the upper bound choice of 1~Gyr. The median and $1\sigma$ dispersion of $\rho$ and $p$-value for the Monte Carlo Spearman's rank correlation are also shown in the bottom left of each panel.
}
\label{fig:incvproperty}
\end{figure*}

The dust emission model fits to the SEDs, from which $L_{\rm TIR}$ is derived, for the four example galaxies can be seen in Figure~\ref{fig:obsmodsed}. The solid blue (\citealt{2000ApJ...533..682C} fits) and orange (inclination-dependent fits) lines represent the best-fit (minimum $\chi^2$) models, with the dust emission dominating beyond $\lambda_{\rm rest} > 5\mu$m. It can be seen that these four galaxies, as well as most other galaxies in the sample, have relatively well-constrained peaks of the dust emission. This is due to the sample requirement of at least one FIR data point being beyond rest frame 100~$\mu$m. However, while it appears from these examples that the dust emission in the MIR and $L_{\rm TIR}$ may vary in agreement between models, that is only for the best-fit values. Since \texttt{Lightning} produces probability distributions for these properties, a better comparison would be of these distributions.

A comparison of the $L_{\rm TIR}$ distributions is displayed in Figure~\ref{fig:incvproperty}(a), which shows the median and 1$\sigma$ difference of the logarithm between the \citet{2000ApJ...533..682C} and inclination-dependent estimates (i.e., logarithm of the ratio) of $L_{\rm TIR}$ versus the inclination derived from the SED fittings. The orange line and corresponding light gray 1$\sigma$ dispersion range are the mean and standard deviation of $\Delta \cos i=0.01$ bins calculated by using all 5000 elements within the MCMC chains of each property as data points (i.e., $58~\rm{galaxies} \times 5000~\rm{chain~elements}=290,000~\rm{data~points}$). From the binned average and data points, it can be seen that the $L_{\rm TIR}$ estimates between models are in excellent agreement for most galaxies, with the average and 1$\sigma$ dispersion being consistent with zero at all inclinations. To illustrate the impact of SFH binning, we show in Figure~\ref{fig:incvproperty} the cases where the upper bound on the third age bin is adjusted to 500~Myr (green lines) and 1.5~Gyr (blue lines) from its original 1~Gy (orange lines). We also computed the Spearman's rank correlation using a Monte Carlo method to check for the presence of any trends between fits. To do this, we selected a random value from the distribution of inclination and $\Delta\log_{10}(L_{\rm TIR})$ for each galaxy and computed Spearman's rank test for the ensemble. This was repeated 5000 times to build up a distribution of $\rho$ and $p$-value, from which to determine the median and 1$\sigma$ values; these are annotated in the bottom left of the panel. For $L_{\rm TIR}$, this shows that we are confident there is no monotonic relation with inclination, and that both fits have similar $L_{\rm TIR}$. Therefore, we concluded that $L_{\rm TIR}$ is comparable between fits for most galaxies and further comparisons between derived properties and their dependence on inclination are meaningful.

Examining the stellar models, the best-fit unattenuated stellar model spectra for both the \citet{2000ApJ...533..682C} and inclination-dependent fits are shown for the four example galaxies as the dotted lines in Figure~\ref{fig:obsmodsed}. It can be seen that the nearly face-on galaxy (i.e., upper most panel) has similar unattenuated spectra. However, for the more inclined galaxies, the unattenuated spectra can vary greatly from the UV to the NIR. This difference is expected due to the significant differences between the attenuation curves at high inclinations, as shown in Figures~\ref{fig:tuffsattenuation} and \ref{fig:tuffsattenuationrold} (i.e., left most panels). This variation in edge-on galaxies is what drives observed differences in some stellar properties between fits with inclination, as shown in Figure~\ref{fig:incvproperty}.

It was expected that the \citet{2000ApJ...533..682C} fits would predict relatively high stellar emission attenuation at low inclinations and relatively low attenuation at high inclinations compared to the inclination-dependent fits. This is because the observed UV-optical flux will vary based on the viewing angle of the galaxy due to inclination-based attenuation, but the dust emission will be nearly unaffected by the viewing angle. Assuming a cylindrical geometry for the disk, we would expect to observe more rest-frame UV-optical flux from a nearly face-on view of a galaxy compared to the average view if it were randomly oriented (i.e., moderately inclined), but would predict similar levels of absorption due to $L_{\rm TIR}$ being unaffected by inclination. As for the edge-on view, we would expect to observe less rest-frame UV-optical flux than the average view, while still predicting similar levels of absorption. Therefore, since an inclination-independent model like the \citet{2000ApJ...533..682C} model should be most applicable to the average galaxy, it would overestimate the line-of-sight attenuation for more face-on galaxies and underestimate the line-of-sight attenuation for edge-on galaxies; contrarily the inclination-dependent model should properly account for inclination-dependent line-of-sight attenuation. 

This effect can indeed be clearly seen in Figure~\ref{fig:incvproperty}(b), which shows the difference in $A_V$ of the two fits versus inclination. From the Spearman's rank correlation, it can be seen that this trend is very strong and highly significant with a median $p\textrm{-value}<10^{-11}$. For face-on to moderately inclined galaxies (i.e., $1-\cos i \lesssim 0.4$) in our $z<0.7$ sample (9 galaxies), not accounting for inclination-based attenuation results in $A_V$ being higher by $0.31\pm0.04$ magnitudes on average, whereas for edge-on galaxies (i.e., $1-\cos i = 0.9$--$1.0$, 14 galaxies) this results in $A_V$ being lower by 0.28--0.67 magnitudes.

However, this expected variation in attenuation with inclination is not seen in the FUV attenuation. Figure~\ref{fig:incvproperty}(c) shows the difference in $A_{\rm FUV}$ with inclination, which has the inclination-dependent model predicting $A_{\rm FUV}\approx0.19$ magnitudes higher on average at all inclinations compared to the \cite{2000ApJ...533..682C} model. The lack of a trend with inclination for our sample could be due to either (1) the inclination-dependent model incorrectly predicting $A_{\rm FUV}$ or (2) the \citet{2000ApJ...533..682C} attenuation model is properly accounting for the inclination-based attenuation at FUV wavelengths at our current levels of uncertainty. To check which occurs, we compared each fit individually with the inclination and found that both fits had a strong increase in $A_{\rm FUV}$ with increasing inclination, which is expected to occur due to inclination-based attenuation. The increase in $A_{\rm FUV}$ with inclination and the expected trend being seen in the difference of $A_V$ led us to conclude that explanation (2) was correct.

The lower average $A_{\rm FUV}$ of the \citet{2000ApJ...533..682C} model for galaxies in our sample leads to a similar lower average of 0.19 dex in SFR$_{100}$ compared to the inclination-dependent model for all inclinations, as seen in Figure~\ref{fig:incvproperty}(d). This is due to SFR$_{100}$ being correlated with $A_{\rm FUV}$ by the young UV emitting stellar population. Like $A_{\rm FUV}$, there is practically no trend with inclination, and because of this relative lack of trend, we conclude that the \citet{2000ApJ...533..682C} attenuation curve can model inclination-based attenuation like the inclination-dependent model at FUV wavelengths and recover the resulting recent SFRs for the disk galaxies in our sample at all inclinations.

The reason for the \citet{2000ApJ...533..682C} model being able to account for the inclination-based attenuation at FUV wavelengths, while also having the expected trend in the $V$-band wavelengths with inclination, can be found in the results from the SED fits. Looking at the correlation between parameters, $A_{\rm FUV}$ for both fits is moderately to strongly correlated with $L_{\rm TIR}$, with Pearson correlation coefficients for a given fit averaged for all galaxies of $\rho=0.40\pm0.13$ and $\rho=0.79\pm0.12$ for the inclination-dependent and \citet{2000ApJ...533..682C} models, respectively. Therefore, since $A_{\rm FUV}$ and $L_{\rm TIR}$ are relatively correlated for a given fit with both models, and $L_{\rm TIR}$ is consistent between models, it results in $A_{\rm FUV}$ being relatively consistent as well. Comparing this to the correlations between $A_V$ and $L_{\rm TIR}$, the inclination-dependent model has a weaker correlation of $\rho=0.32\pm0.16$, while the \citet{2000ApJ...533..682C} model still has a strong correlation of $\rho=0.79\pm0.12$. The strong correlation for both $A_{\rm FUV}$ and $A_V$ with $L_{\rm TIR}$ for the \citet{2000ApJ...533..682C} model is due to the use of a single normalized attenuation curve, which causes a correlation of 1 between $A_{\rm FUV}$ and $A_V$. However, $A_V$ has practically no correlation with $A_{\rm FUV}$ ($\rho=0.01\pm0.24$) for a given fit with the inclination-dependent model, which allows for the expected trend with inclination and the difference between models. Thus, $A_{\rm FUV}$ and $A_V$ are controlled by $L_{\rm TIR}$ for the \citet{2000ApJ...533..682C} model, whereas only $A_{\rm FUV}$ is controlled by $L_{\rm TIR}$ for the inclination-dependent model, and $A_V$ can be a variety of values for a given $A_{\rm FUV}$.

This trend in the difference of $A_V$ with inclination is also seen in the stellar mass logarithmic differences, due to the intrinsic optical emission, which $A_V$ represents, dominating the stellar masses estimates. The logarithmic difference in $M_\star$ with inclination can be seen in Figure~\ref{fig:incvproperty}(e), which shows a moderate, statistically significant trend with a median $p\textrm{-value}<10^{-2}$. From the panel for stellar mass, it can be seen that the mass is relatively consistent on average between fits with the \citet{2000ApJ...533..682C} fits producing a slightly lower estimate of $M_\star$ compared to the inclination-dependent fits by a factor of $-0.05\pm0.03$ dex over $1-\cos i = 0$--$0.6$ (19 galaxies). However, at $1-\cos i \gtrsim 0.7$ (36 galaxies), the \citet{2000ApJ...533..682C} fits produce lower estimates of $M_\star$ compared to the inclination-dependent fits. Not including inclination-based attenuation can lead to lower $M_\star$ values by a factor of $0.12\pm0.12$ dex at $1-\cos i \approx 0.75$ up to $0.17\pm0.15$ dex at $\approx90^{\circ}$ for galaxies in our $z<0.7$ sample. This result is consistent with the findings from \citet{2007MNRAS.379.1022D} and \citet{2018MNRAS.480.3788W} who found stellar mass estimates (from inclination corrected mass-to-light ratios) to be inclination-independent for $i \lesssim 70^{\circ}$. However, above that angle, they found that stellar masses may be underestimated by a factor of $\approx$0.3 dex for an inclination-independent model.
 
Finally, Figure~\ref{fig:incvproperty}(f) shows the logarithmic difference in sSFR$_{100}$ with inclination. Since sSFR$_{100}$ is computed as SFR$_{100}$ divided by the $M_\star$, the trend seen with inclination is a combination of the trend seen in SFR$_{100}$ and the reflected trend seen in the stellar mass. This trend is not nearly as strong or statistically significant compared to the stellar mass due to the large dispersion and uncertainties introduced by SFR$_{100}$. Overall, sSFR$_{100}$ is somewhat lower for the \citet{2000ApJ...533..682C} fits by a factor of 0.14 dex from $1-\cos i = 0$--$0.8$ transitioning to becoming larger by a factor of 0.05 dex at 90$^{\circ}$.

\section{Summary} \label{sec:Summary}

We developed and tested an inclination-dependent attenuation module for the SED fitting code \texttt{Lightning}, in order to test the effects of inclination-based attenuation on derived SFHs. The module utilizes the inclination-dependent attenuation curve from \citet{2004A&A...419..821T} as updated by \citet{2011A&A...527A.109P}. We tested the module using 82 disk-dominated galaxies, as determined by their S\'ersic index ($n<1.2$) and subsequent visual inspection, that had UV to FIR data from the GOODS North and South fields.

Using the measured axis ratio $q$ of each galaxy from \citet{2012ApJS..203...24V}, we derived PDFs of inclination from a Monte Carlo method that incorporates the distributions of the intrinsic thickness and asymmetry of spiral galaxies from \citet{2013MNRAS.434.2153R}. We found that these PDFs give median inclinations that are in excellent agreement with inclinations that are derived from Equation~\ref{eq:cosi} with commonly used fixed values of $\gamma$ if $i \gtrsim 30^{\circ}$. However, the inclination uncertainties derived from Equation~\ref{eq:cosi} for all inclinations are generally underestimated by a factor of $\approx$7.5 compared to our inclination PDFs.

We then fitted the SEDs of our sample galaxies twice, first with the inclination-independent \citet{2000ApJ...533..682C} attenuation curve, and second with an inclination-dependent attenuation model. In order to accurately model the SEDs with the inclination-dependent model, we found that prior distributions on the inclination were required. With the priors, most inclinations ($\approx$88\%) derived from the SED fits when compared to the inclination PDFs derived from the measured $q$ had intersection-area-to-union-area ratios $>0.05$, which we considered in agreement. Those that did not tended to have elevated $L_{\rm TIR}$ and higher redshifts ($z\gtrsim0.7$). It is possible that this is due to the \citet{2004A&A...419..821T} attenuation curves not being physically appropriate to model most galaxies above this redshift, because galaxies with $z\gtrsim1$ tend to be significantly thicker and dynamically hotter than galaxies with $z<1$ \citep[e.g.,][]{2013ApJ...773...43B,2014ApJ...792L...6V,2017ApJ...847...14E,2019MNRAS.490.3196P,2019MNRAS.484.5170Z}.

Limiting the 82 galaxy sample to only include the 58 galaxies with $z<0.7$ as to mitigate any redshift evolution effects, we compared the inclination-dependent and \citet{2000ApJ...533..682C} fits for this $z<0.7$ sample. We found that both fits recover the expected trend with inclination for $A_{\rm FUV}$ and average SFRs of the last 100 Myr at all inclinations. By contrast, not accounting for inclination-based attenuation in our sample of galaxies resulted in an average $A_V$ being elevated by $0.31\pm0.04$ magnitudes for face-on to moderately inclined galaxies and underestimated by 0.28--0.67 magnitude for edge-on galaxies. Stellar masses were in good agreement between fits for $1-\cos i = 0$--$0.6$ with a minor scatter of $\approx$0.1 dex. For $1-\cos i \gtrsim 0.75$, stellar masses could be underestimated up to a factor of 0.17 dex at 90$^{\circ}$ by the \citet{2000ApJ...533..682C} model compared to the inclination-dependent model. These results indicated for our sample of galaxies that the \citet{2000ApJ...533..682C} attenuation curve is able to correctly model the inclination-dependent attenuation of FUV emission, which dictates the recent SFRs, at all inclinations; but fails for the optical-NIR emission, which dominates the stellar masses estimates, at higher inclinations.

This work introduces and shows the impact of inclination-dependent attenuation on SFHs, and subsequently stellar masses and recent SFRs, derived from SED fitting. Incorporating inclination-dependent attenuation when fitting SEDs can help give better insight into the physical properties of highly inclined galaxies. In an upcoming paper, we use this inclination-dependent model to determine how inclination affects $A_{\rm FUV}$ calibrations that are used to compute SFRs and compare the results with previously published works. Beyond this, we intend to apply the inclination-dependent attenuation module to galaxies that have sizable bulge components, and a more complete sample of galaxies to test whether our results hold for the broader disk-galaxy population.

\acknowledgments

{We acknowledge and thank the anonymous referees for their valuable and insightful comments, which significantly helped improve the quality of this paper. We gratefully acknowledge support from the NASA Astrophysics Data Analysis Program (ADAP) grant 80NSSC20K0444 (KD, RTE, BDL). The material in this paper is based upon work supported by NASA under award number 80GSFC21M0002. This work is based on observations taken by the CANDELS Multi-Cycle Treasury Program with the NASA/ESA \textit{HST}, which is operated by the Association of Universities for Research in Astronomy, Inc., under NASA contract NAS5-26555. This work has made use of the Rainbow Cosmological Surveys Database, which is operated by the Centro de Astrobiolog\'ia (CAB/INTA), partnered with the University of California Observatories at Santa Cruz (UCO/Lick, UCSC); the NASA/IPAC Extragalactic Database (NED), which is funded by the National Aeronautics and Space Administration and operated by the California Institute of Technology; and the Arkansas High Performance Computing Center, which is funded through multiple National Science Foundation grants and the Arkansas Economic Development Commission.}

\facilities{HST, Spitzer, Herschel, Blanco, VLT:Melipal, VLT:Yepun, Mayall, LBT, Subaru, CFHT}

\software{\texttt{Lightning} (Eufrasio et al. 2017), \texttt{P\'EGASE} (Fioc \& Rocca-Volmerange 1997, 1999)}

\clearpage

\bibliographystyle{aasjournal}
\bibliography{Inclination}

\appendix

\section{Mid-to-Far IR Photometry Assessment} \label{sec:IRAssess}

Given the relatively large Herschel FIR PSFs, we tested the potential impact of blending and/or background fluctuations \citep{2011A&A...533A.119E,2013A&A...553A.132M} that may be present in our final sample.  As noted by \citet{2019ApJS..243...22B}, there should be minimal confusion of source identifications due to their procedure of using the higher resolution MIPS 24~$\mu$m source locations as positional priors when determining the PACS and SPIRE counterparts. However, photometric issues could potentially arise due to nearby IR-bright sources.

We first visually inspected the PACS and SPIRE images for any obvious PSF blending, at the locations of our MIPS 24~$\mu$m sources that could impact the FIR photometry.  We found negligible bright-source PSF overlap for all PACS bands and the SPIRE 250~$\mu$m band.  However, the SPIRE 350 and 500~$\mu$m sources showed nonnegligible PSF overlap, and we therefore chose to exclude photometry based on these two bands when fitting the SEDs.

We further assessed whether the remaining FIR photometry was reliable, and did not suffer from important photometric blending from multiple bright sources within the PSFs. Using the MIPS 24~$\mu$m counterpart flags in Table 18 of \citet{2019ApJS..243...22B}, we determined the number of MIPS 24~$\mu$m counterparts within each of the PACS and SPIRE 250~$\mu$m band PSFs and the contributions of the primary source counterpart to the full 24~$\mu$m flux within the PSFs. Using a PSF with a FWHM of 7$^{\prime\prime}$, 11.2$^{\prime\prime}$, and 18$^{\prime\prime}$ for the PACS 100~$\mu$m, 160~$\mu$m, and SPIRE 250~$\mu$m, respectively, we found 5\%, 18\%, and 45\% of our final sample that had the respective band contained more than one 24~$\mu$m detected source within the FWHM diameters. This indicated that the 160~$\mu$m and 250~$\mu$m bands may have some nonnegligible source confusion.  However, for the 160~$\mu$m and 250~$\mu$m bands, 93\% and 62\% of the respective sources with more than one 24~$\mu$m counterpart had 24~$\mu$m fluxes dominated by the primary counterpart. This implies that the majority of the 160~$\mu$m and 250~$\mu$m sources with potentially blended counterparts would have fluxes elevated by $< 100$\%, with a median elevation of 41\% and 39\%, respectively, if all counterpart sources are blended. This minimal level of blending was expected, since \citet{2019ApJS..243...22B} found their mid-to-far IR photometry in the GOODS-N was in excellent agreement with the superdeblended photometry in \citet{2018ApJ...853..172L}.

In addition to source blending, low signal-to-noise ratio FIR photometry may be impacted by fluctuations in local backgrounds. \citet{2011A&A...533A.119E} and \citet{2013A&A...553A.132M} used Monte Carlo simulations and showed that, at the $3\sigma$ limit adopted for our sample selection, the photometric accuracy is better than 33\% for at least 68\% of sources. This accuracy improves with increasing signal-to-noise ratio. Given that all FIR sources in our sample have signal-to-noise ratios~$>3$ and the majority of PACS sources (72\% and 93\% for PACS 100~$\mu$m and 160~$\mu$m, respectively) have signal-to-noise ratios~$>5$, we expect minimal photometric issues from background fluctuations in the PACS bands and expect the photometry to be highly accurate. We find that 37\% of SPIRE 250~$\mu$m sources have signal-to-noise ratios~$>5$, so it is possible that the SPIRE 250~$\mu$m photometry may suffer from lower accuracy due to this confusion.

Since a sizable fraction of the SPIRE 250~$\mu$m sources in our sample could have nonnegligible blending and background fluctuations, we investigated the effects of including the SPIRE 250~$\mu$m photometry on the SED fits by refitting the 67 galaxies in the final sample that had the SPIRE 250~$\mu$m band without the SPIRE 250~$\mu$m band. We found that sources that had PACS 160~$\mu$m to constrain the peak of the dust emission had bolometric luminosities for the dust emission models always within 19\% of the luminosities including the SPIRE 250~$\mu$m band, with  a median and scatter of $\log(L_{\rm IR}^{\rm with}/L_{\rm IR}^{\rm without}) = 0.00\pm0.03$ for the sample, where $L_{\rm IR}^{\rm with}$ and $L_{\rm IR}^{\rm without}$ are the bolometric luminosities from the fits with and without the SPIRE 250~$\mu$m band, respectively. However, sources that did not have the PACS 160~$\mu$m to constrain the peak of the dust emission could have bolometric luminosities that vary up to 45\% from the luminosities including the SPIRE 250~$\mu$m band, with $\log(L_{\rm IR}^{\rm with}/L_{\rm IR}^{\rm without}) = 0.03\pm0.07$. Therefore, we utilize the the SPIRE 250~$\mu$m band in our fits due to it having minimal adverse effects on these fits and the beneficial effect of helping constrain the peak of dust emission.

\section{Diagnostic Figures} \label{sec:DiagFigs}

\begin{figure*}[h!]
\centerline{
\includegraphics[width=17cm]{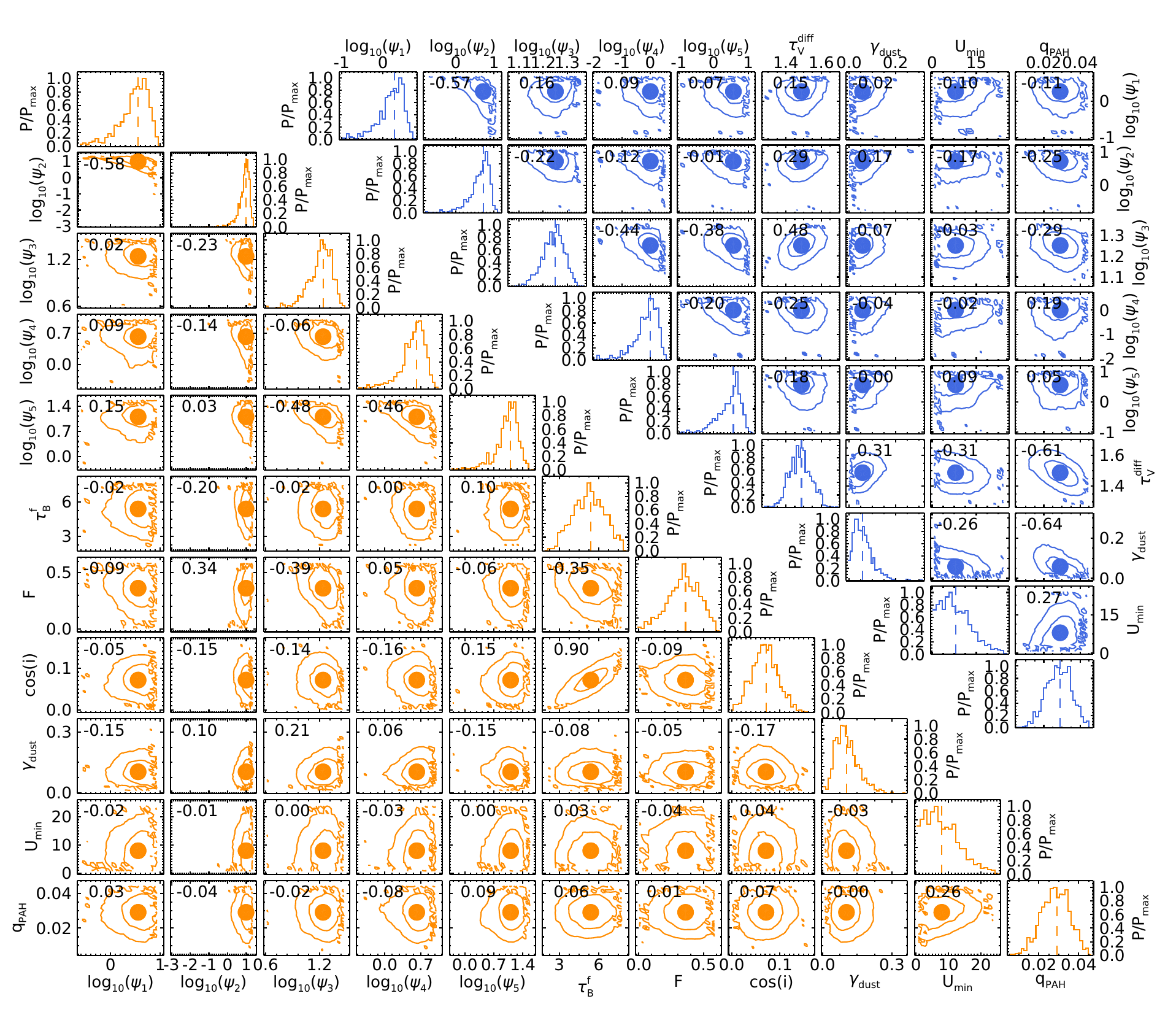}
}
\caption{
(\textit{Lower left triangle plot}): Probability distribution functions in terms of $P/P_{\rm max}$ (diagonal elements) and the 68\% and 95\% confidence contours for the free parameter pairs (off-diagonal elements) in our inclination-dependent fits with the image-based inclination prior for our most inclined example galaxy, J123555.43+621056.8. This galaxy is the galaxy with a purple outline in Figure~\ref{fig:postagestamp}. The vertical dashed lines in the histograms and solid colored circles in the contour plots indicate the median values of each parameter. The Pearson correlation coefficients for each set of parameters are shown in the upper left corners of each contour plot. (\textit{Upper right triangle plot}): Same as the lower left, except for the free parameters in the \citet{2000ApJ...533..682C} fits (see Table~\ref{table:LightParam} for a list of parameters and units).
}
\label{fig:tridistoutparam}
\end{figure*}

\begin{figure*}[t!]
\centerline{
\includegraphics[width=18cm]{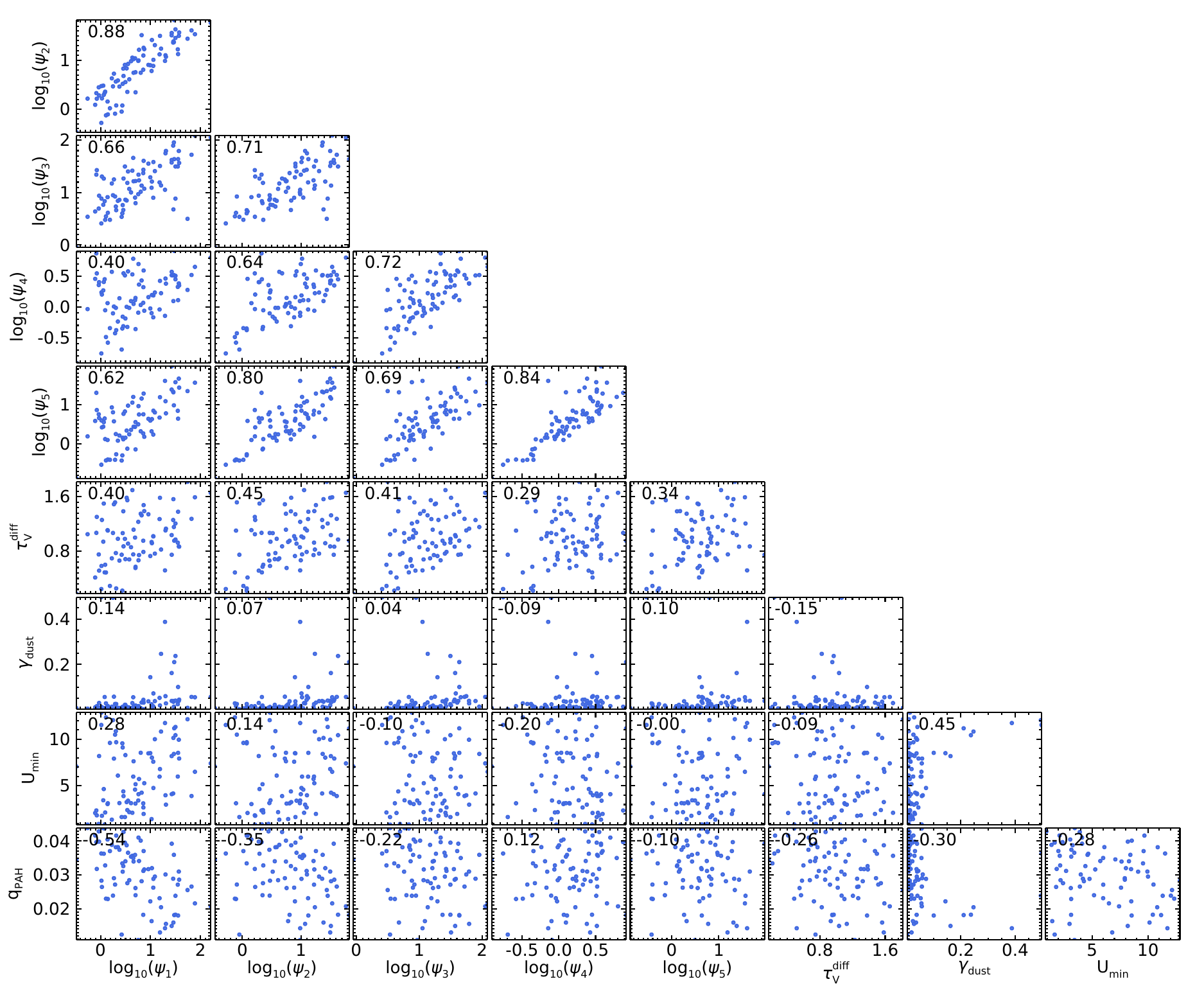}
}
\caption{
Scatter plots for the free parameter pairs in our \citet{2000ApJ...533..682C} fits. Each point represents the median value of that parameter for a galaxy in our sample. The Pearson correlation coefficients for each set of parameters are shown in the upper-left corners of each scatter plot. These coefficients show the global trends in the data. Most correlations seen were expected, such as that between $\psi_i$ and $\tau_V^{\rm diff}$, and $\psi_i$ and $\psi_j$, where $i$ and $j$ are different age bins. These strong correlations between $\psi_i$ and $\tau_V^{\rm diff}$ are due to the increased attenuation allowing for larger SFRs (See Table~\ref{table:LightParam} for list of parameters and units).
}
\label{fig:diagscattercal}
\end{figure*}

\begin{figure*}[t!]
\centerline{
\includegraphics[width=18cm]{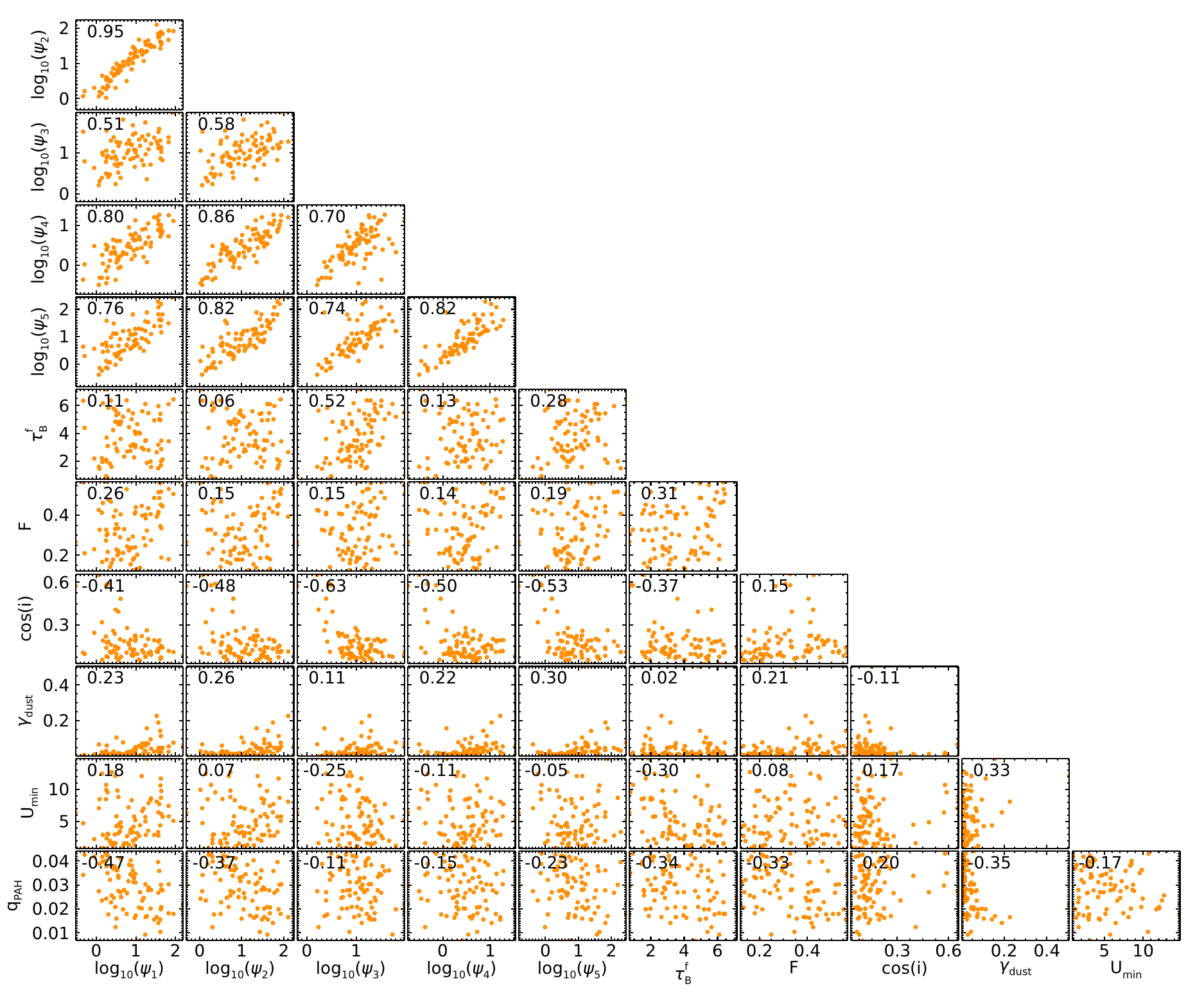}
}
\caption{
Same as Figure~\ref{fig:diagscattercal} except for the inclination-dependent fits with a flat inclination prior.
}
\label{fig:diagscattertuffsnp}
\end{figure*}

\begin{figure*}[t!]
\centerline{
\includegraphics[width=18cm]{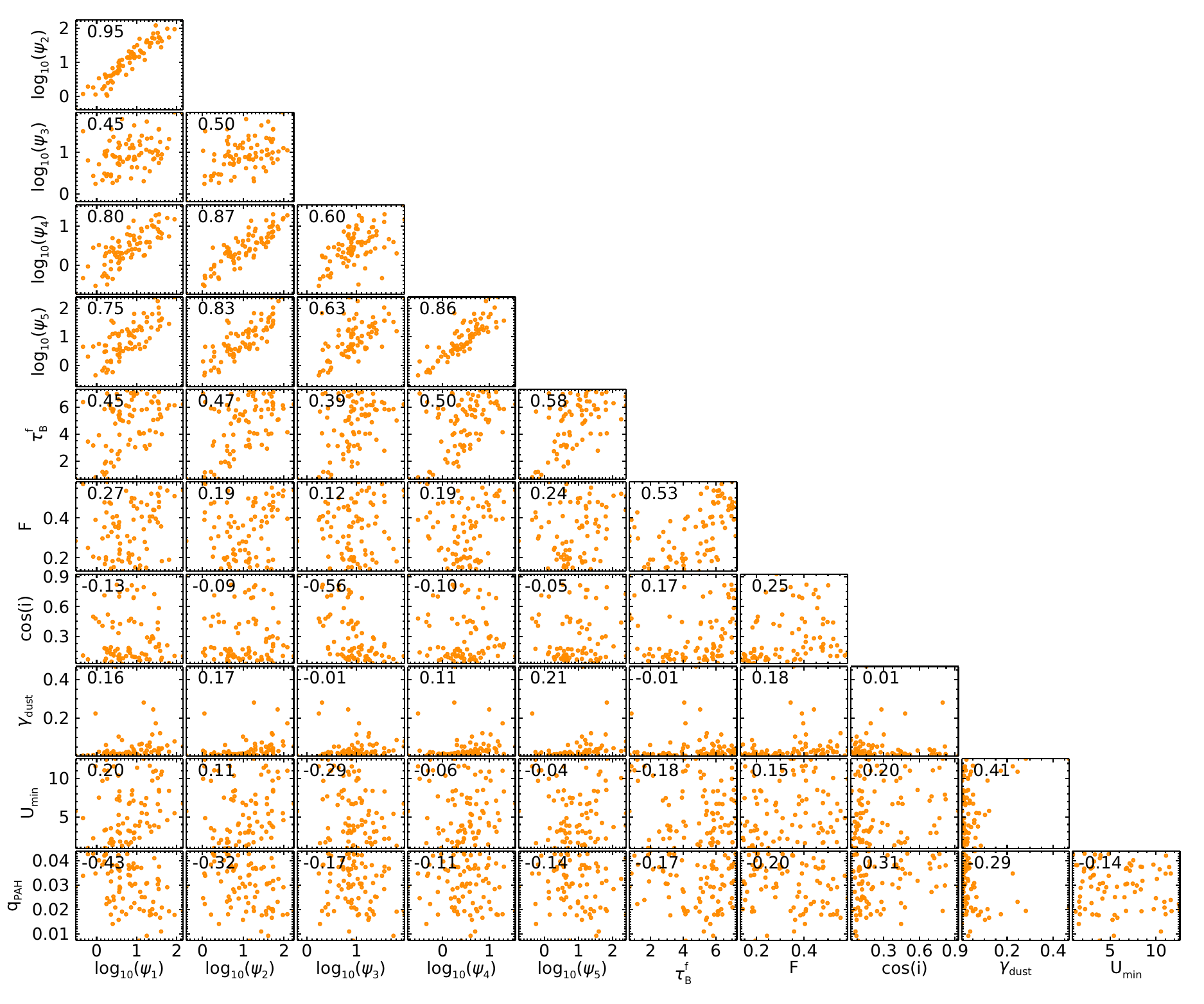}
}
\caption{
Same as Figure~\ref{fig:diagscattercal} except for the inclination-dependent fits with the image-based inclination prior. Like the \citet{2000ApJ...533..682C} fits, we see the expected correlations between $\tau_B^f$ and $\psi_i$, and $\psi_i$ and $\psi_j$, where $i$ and $j$ are different age bins. One other notable feature is the effect of using the image-based inclination prior on the inclination-$\tau_B^f$ degeneracy. As seen in Figure~\ref{fig:diagscattertuffsnp}, $\cos i$ and $\tau_B^f$ have a slight correlation. However, this correlation is minimized after implementing the image-based prior, implying that using this prior helps mitigate the inclination-$\tau_B^f$ degeneracy.
}
\label{fig:diagscattertuffs}
\end{figure*}

\begin{figure*}[t!]
\centerline{
\includegraphics[width=18cm]{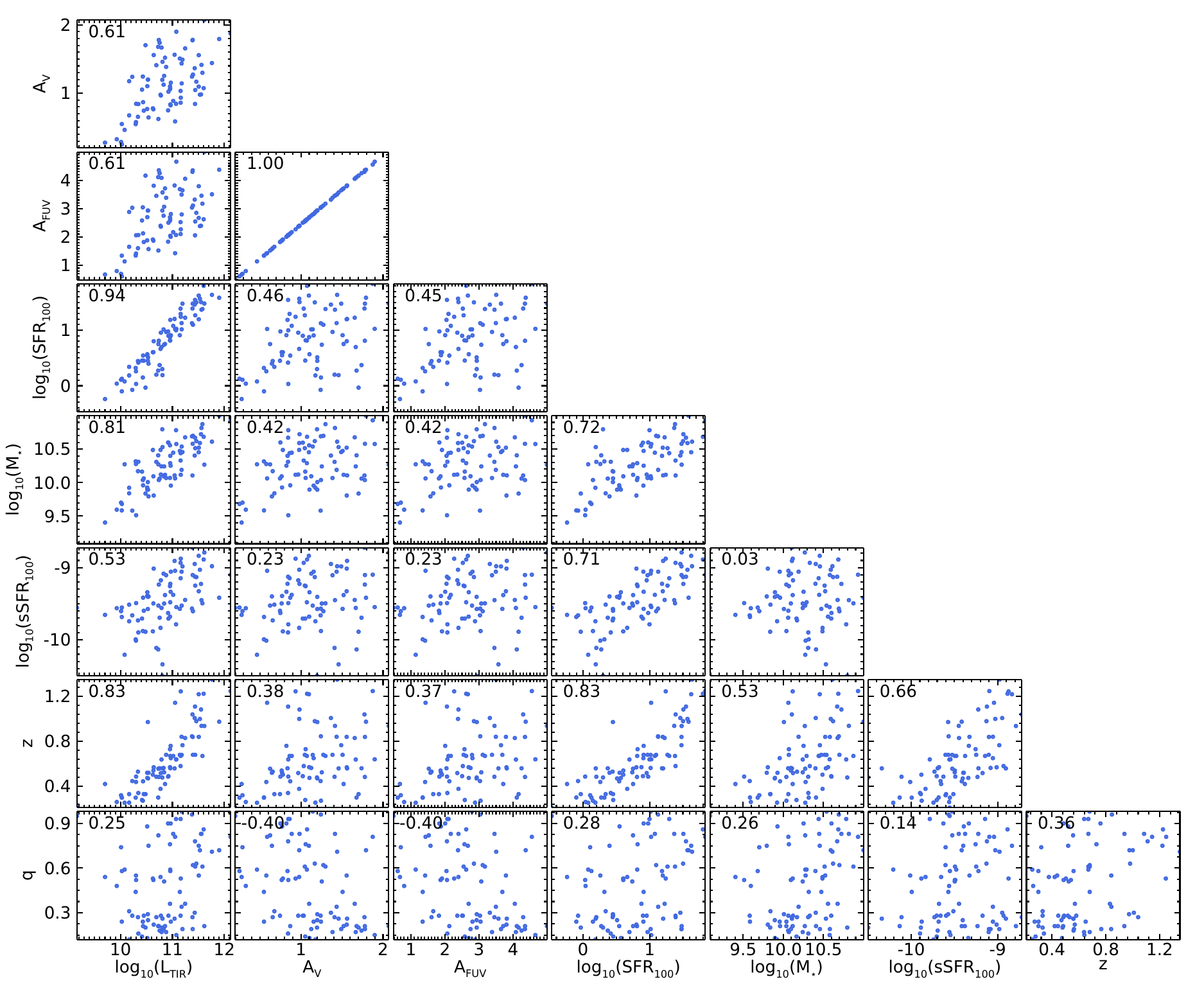}
}
\caption{
Scatter plots for the parameter of interest ($L_{\rm{TIR}} \ [L_\odot]$ , $A_V \ [{\rm mag}]$, $A_{FUV} \ [{\rm mag}]$, $\rm{SFR}_{100} \ [M_\odot \ yr^{-1}]$, $M_{\star} \ [M_\odot]$, and $\rm{sSFR}_{100} \ [yr^{-1}]$) of our \citet{2000ApJ...533..682C} fits, along with redshift $z$ and axis ratio $q$. Each point represents the median or given value of that parameter for a galaxy in our sample. The Pearson correlation coefficients for each set of parameters are shown in the upper-left corners of each scatter plot and give the global trends in the data. It is important to stress that these correlations are for the global trends and not the average of the individual fits as in Section~\ref{sec:Discussion}. Most correlations seen were expected, such as that between $L_{\rm TIR}$ and all other properties besides $q$. These positive correlations with $L_{\rm TIR}$ are due to the energy balance assumption which requires larger attenuation and SFRs with increasing $L_{\rm TIR}$. As for $z$ and $L_{\rm TIR}$, this correlation is a direct result of our selection process, which would require brighter IR emission at higher redshifts in order for the galaxy to be detected. It is also important to note that all parameters, besides $A_V$ and $A_{FUV}$, are relatively independent of $q$, which confirms that selection effects are not significantly biasing our sample such that our results in Section~\ref{sec:Discussion} would be influenced by this bias.
}
\label{fig:diagparintcal}
\end{figure*}

\begin{figure*}[t!]
\centerline{
\includegraphics[width=18cm]{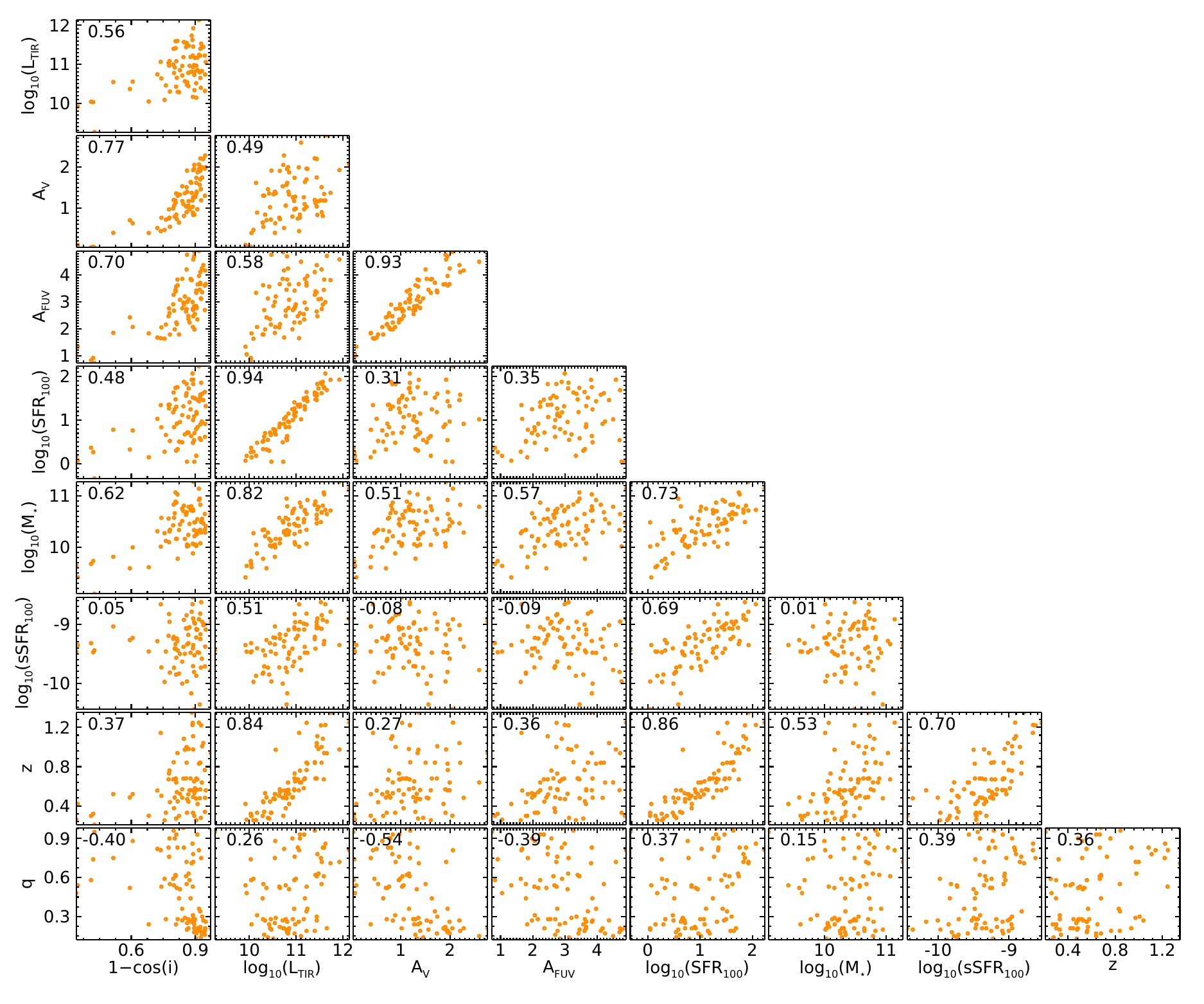}
}
\caption{
Same as Figure~\ref{fig:diagparintcal} except for the inclination-dependent fits with a flat inclination prior and the addition of the inclination parameter ($1-\cos i$).
}
\label{fig:diagparinttuffsnp}
\end{figure*}

\begin{figure*}[t!]
\centerline{
\includegraphics[width=18cm]{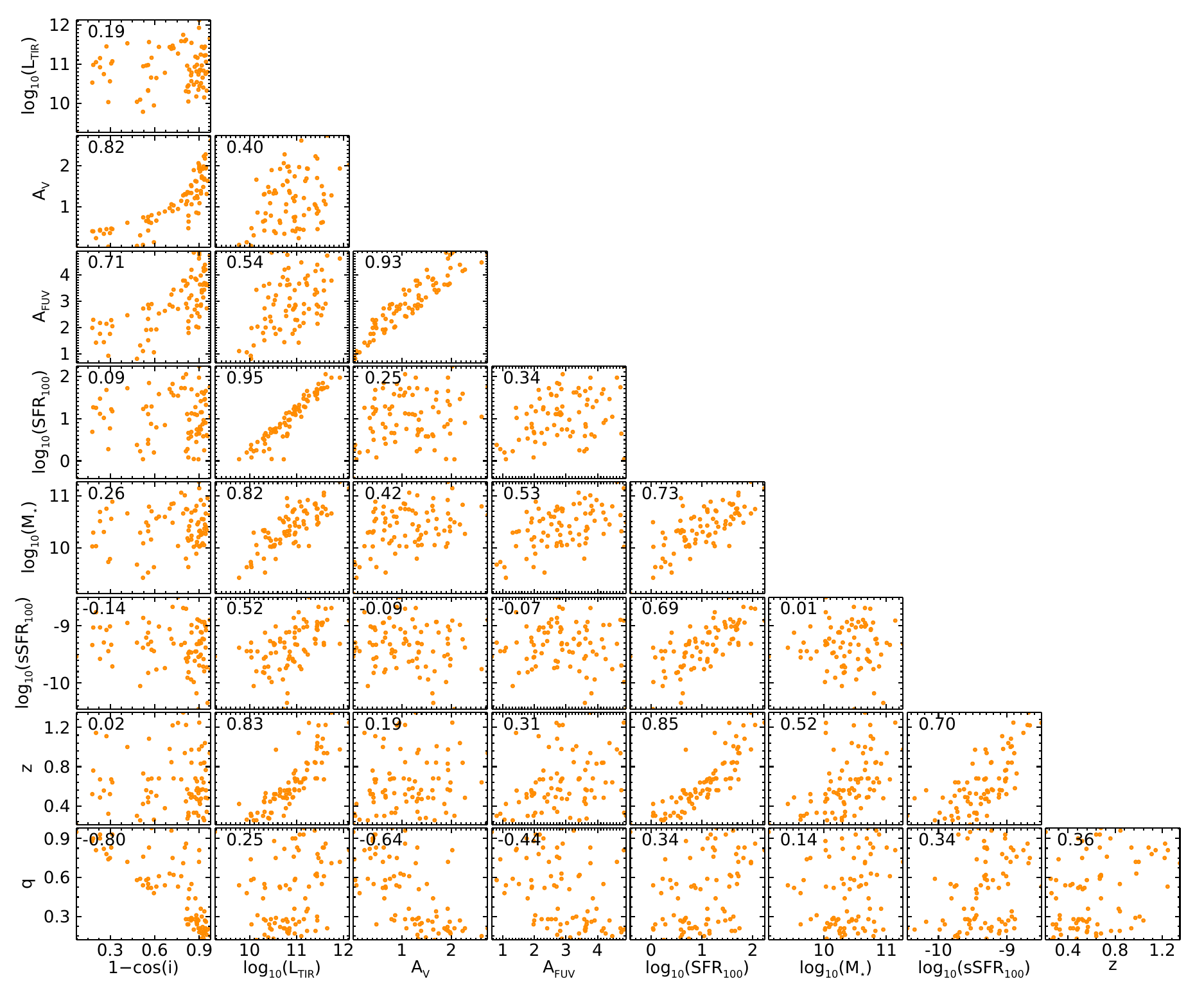}
}
\caption{
Same as Figure~\ref{fig:diagparintcal} except for the inclination-dependent fits with the image-based inclination prior and the addition of the inclination parameter ($1-\cos i$). Like the \citet{2000ApJ...533..682C} fits, we see the expected correlations between $L_{\rm TIR}$ and the other parameters besides $q$ and inclination. Again, minimal correlations can be seen between parameters (excluding $A_V$ and $A_{\rm FUV}$) and $q$, which further implies that selection effects are not significantly biasing our sample. One other notable feature is the correlation between inclination and $q$ with attenuation, which shows that more inclined galaxies tend to have increased attenuation.
}
\label{fig:diagparinttuffs}
\end{figure*}

To show any degeneracies between parameters and their location in parameter space, diagnostic plots showing free parameter distributions and the global trends for all galaxies are provided. Figure~\ref{fig:tridistoutparam} shows the distributions for the free parameters in the \citet{2000ApJ...533..682C} fit and inclination-dependent fit with an image-based inclination prior for our most inclined example galaxy, J123555.43+621056.8, in the upper right and lower left, respectively. Figures~\ref{fig:diagscattercal}--\ref{fig:diagscattertuffs}  show scatter plots of the median of each galaxies' free parameter pairs to display the global trends for these parameters for the \citet{2000ApJ...533..682C} fits, inclination-dependent fits with a flat inclination prior, and inclination-dependent fits with an image-based inclination prior, respectively. Figures~\ref{fig:diagparintcal}--\ref{fig:diagparinttuffs} also show median parameter scatter plots, but for the parameters of interest and additionally redshift and axis ratio.

\end{document}